\RequirePackage{fix-cm}
\documentclass[smallextended]{svjour3}       
\smartqed  
\usepackage{graphicx}
\usepackage{cite}
\usepackage{amsmath,amssymb,amsfonts}
\usepackage[ruled,linesnumbered]{algorithm2e}
\usepackage{algorithmic}
\usepackage{graphicx}
\usepackage{url}
\usepackage{soul}
\usepackage{textcomp}
\usepackage[table]{xcolor}
\usepackage{mwe,graphicx,tabu,subcaption,float}
\DeclareMathOperator*{\argmax}{arg\,max}
\SetKwComment{Comment}{$\triangleright$\ }{}
\SetKwInOut{Parameter}{Parameters}
\begin{document}

\title{Embedding-based Silhouette Community Detection}

\titlerunning{Silhouette community detection}        

\author{Bla\v{z} \v{S}krlj \and Jan Kralj  \and Nada Lavra\v{c}}


\institute{Bla\v{z} \v{S}krlj \at
             Jo\v{z}ef Stefan Institute, Jamova 39, 1000 Ljubljana, Slovenia \\Jo\v{z}ef Stefan Int. Postgraduate School, Jamova 39, 1000 Ljubljana, Slovenia \\ \email{blaz.skrlj@ijs.si} \\
             Jan Kralj \at
             Jo\v{z}ef Stefan Institute, Jamova 39, 1000 Ljubljana, Slovenia \\
             \email{jan.kralj@ijs.si}\\
             Nada Lavra\v{c}\\
             Jo\v{z}ef Stefan Institute, Jamova 39, 1000 Ljubljana, Slovenia \\                          University of Nova Gorica, Vipavska 13, 5000 Nova Gorica, Slovenia \\
             \email{nada.lavrac@ijs.si}
             }
             
\date{Received: date / Accepted: date (NOTE: this is a preprint)}

\maketitle

\begin{abstract}
Mining complex data in the form of networks is of increasing interest in many scientific disciplines. Network communities correspond to densely connected subnetworks, and often represent key functional parts of real-world systems. In this work, we propose Silhouette Community Detection (SCD), an approach for detecting communities, based on clustering of network node embeddings, i.e. real valued representations of nodes derived from their neighborhoods.
We investigate the performance of the proposed SCD approach on 234 synthetic networks, as well as on a real-life social network. Even though SCD is not based on any form of modularity optimization, it performs comparably or better than state-of-the-art community detection algorithms, such as the InfoMap and Louvain algorithms.
Further, we demonstrate how SCD's outputs can be used along with domain ontologies in semantic subgroup discovery, yielding human-understandable explanations of communities detected in a real-life protein interaction network.
Being embedding-based, SCD is widely applicable and can be tested out-of-the-box as part of many existing network learning and exploration pipelines.

\end{abstract}

\section{Introduction}

Mining complex data in the form of networks is of increasing interest in many scientific disciplines: social, biological, manufacturing and similar systems can be represented and analyzed using network-based approaches. In this work we explore whether the state-of-the-art machine learning techniques for representation learning can effectively be used in unsupervised network analysis.

Recently developed embeddings technology offers advancements in representation learning \cite{zhang2018survey} from different data formats, including learning representations of network data, such as  network node embeddings \cite{cai2018comprehensive}.  Even though such embeddings are commonly used for supervised learning, such as node classification and link prediction, less attention is devoted to the study of how the latent \emph{organization} of a network can be automatically extracted from node embeddings in an \emph{unsupervised} manner.

Real-world complex networks are commonly investigated in terms of their meso-scale topological structure, such as communities \cite{harenberg2014community}. Algorithms such as InfoMap \cite{rosvall2009map,Schaub2017} and Louvain algorithm \cite{de2011generalized} are well established for the task of \emph{community detection}. Identification of communities for example offers insights into the inner workings of cellular function, mobile and transportation networks, and helps with the identification of potential security threats.

\begin{figure}[b!]
\centering
\includegraphics[width=.8\linewidth]{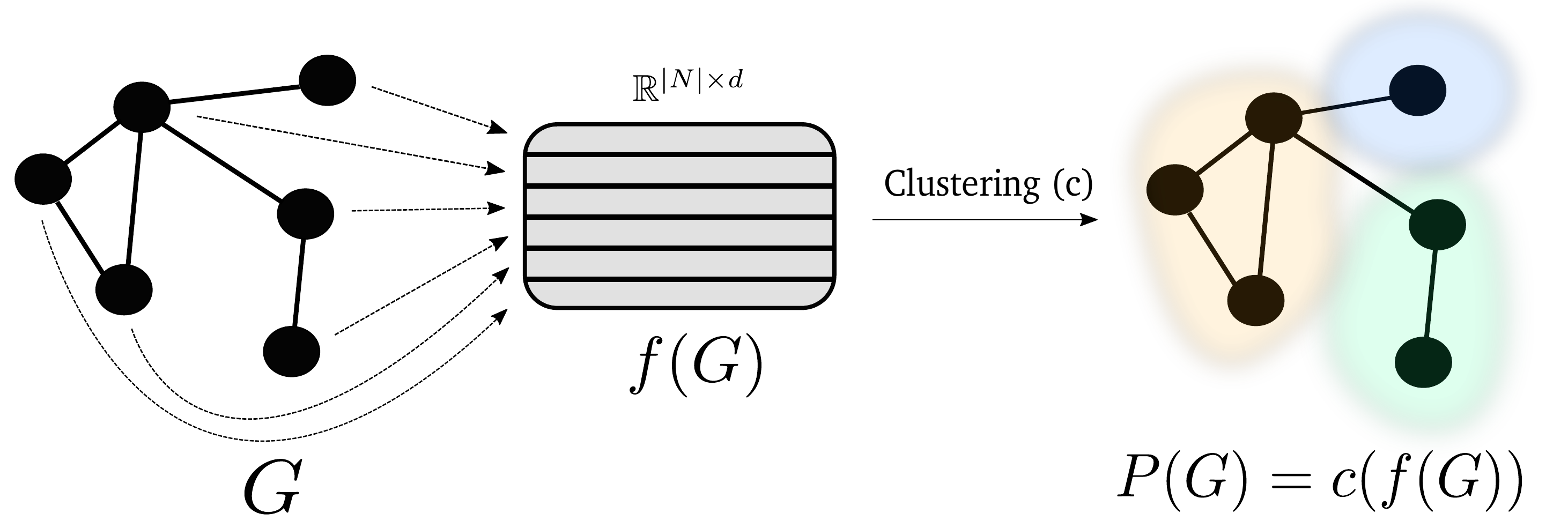}
\caption{Schematic representation of the proposed approach. The input network $G$ (comprised of a set of nodes $N$) first gets mapped to a latent $d$-dimensional vector space via $f$. The vectors are clustered (via $c$) to obtain the final partition $P(G)$.}
\label{fig:scheme}
\end{figure}

The goal of this work is to bridge the two domains, network embedding and community detection, by demonstrating that node embeddings, when clustered, offer insights into network community structure.
Being able to detect communities from embeddings directly could greatly reduce the complexity of existing computational pipelines, comprised of multiple different methods, as well as speed up the development process. The purpose of this work is thus to explore whether node embeddings can be used to build a scalable and fast community detection algorithm. To achieve that, we apply a geometric measure of partition quality (the Silhouette score) on node embeddings directly. The proposed approach is summarized in Figure~\ref{fig:scheme}.

The contribution of this work is multifold, and can be summarized as follows:
\begin{enumerate}
\item We show how the community detection problem can be cast as a network embedding clustering task via optimization of the Silhouette metric.
\item The method is evaluated against strong, established baselines on a large synthetic network collection comprised of 234 networks, where it offers competitive performance. 
\item It is also evaluated on a real-world network, comprised of E-mail communities, where it outperforms the state-of-the-art.
\item The proposed method scales to large networks and can be used with arbitrary node representation learners.
\item We re-implemented NetMF (one of the embedding methods used) in PyTorch \cite{paszke2017automatic} for maximum efficiency and for improved scalability. The method runs on GPGPUs, as well as in parallel on CPUs.
\item On a real-life biological network we showcase how the communities detected with SCD can be \emph{interpteted} via external background knowledge in the form of ontologies, yielding simple rules which describe (in a human-understandable way) the detected communities.
\end{enumerate}

In the following sections, we start with the related work in Section~\ref{sec-RelatedWork}. In the central section of this paper we present the proposed methodology (Section~\ref{sec:scd}). We describe the empirical evaluation of the proposed methodology in Section~\ref{sec:experimental}, followed by the results related to community detection capabilities of SCD (Section~\ref{sec:results}). This is followed by presenting a method enabling the interpretation of the obtained communities (Section~\ref{sec:explain}). We finally discuss the obtained results along with further work (Section~\ref{sec:discussion}).

\section{Related work}
\label{sec-RelatedWork}

In this section we discuss recent attempts at community detection from node embeddings. We begin by
describing state-of-the-art methods for network embedding (Section~\ref{sec-EmbeddingRelatedWork}), followed by the presentation of community detection methods (Section~\ref{sec-CommunityDetectionRelatedWork}). This two methodologies form the basis of the proposed embedding-based Silhouette community detection algorithm (SCD), presented in Section~\ref{sec:scd}.  The communities discovered by SCD are in Section~\ref{sec:explain} explained using semantic subgroup methodology, briefly outlined at the end of this section (Section~\ref{sec-SSDRelatedWork}).

\subsection{Network embedding}
\label{sec-EmbeddingRelatedWork}

A series of the recently introduced methods from the field of representation learning attempt to learn from complex networks by first transforming them into a vectorized form, and then performing a desired down-stream learning task. Such \emph{embeddings} (real-valued, vector representations of nodes) are for example useful for construction of large-scale recommender systems and similar tasks \cite{zhao2016learning}. Some of the well known node embedding algorithms include DeepWalk \cite{perozzi2014deepwalk}, its extension node2vec \cite{grover2016node2vec}, LINE \cite{tang2015line}, struc2vec \cite{ribeiro2017struc2vec} and PTE \cite{tang2015pte}. The common property of these algorithms is that they perform the so-called ``shallow embedding'', exploiting only a given network's adjacency structure. On the other hand, apart from the adjacency structure, methodology from the growing field of geometric deep learning attempts to also exploit various node or edge attributes, present in real-world networks. Intensive development of such methods started with the recently introduced graph convolutional neural networks \cite{defferrard2016convolutional}, as well as for example the graph attention networks \cite{velivckovic2017graph}, graphSAGE \cite{hamilton2017inductive} and many others, recently summarized in \cite{cai2018comprehensive,wu2019comprehensive}.

The common point to all of the aforementioned methods is, they mostly produce real-valued matrices, representing various aspects of a network, let that be the embeddings of desired subnetworks or just network nodes. Such vectorized form of e.g., nodes is a suitable input for the body of well established unsupervised pattern detection methods, such as for example the k-means \cite{bachem2016fast} or K-medoids families \cite{park2009simple} of algorithms. Such clustering methodology has been in widespread use in the statistics and machine learning communities since the 1960s.

Cluster quality can be evaluated with several measures, including Davies-Bouldin Index \cite{bouldin}, Calinski-Harabasz Index \cite{calinski}, Fowlkes-Mallows scores \cite{fowlkes1983method} and the Silhouette score \cite{rousseeuw1987silhouettes}. The latter, used in this work, is 
described in detail in Section~\ref{sec-SilhouetteScore}.

\subsection{Community detection}
\label{sec-CommunityDetectionRelatedWork}

The field of community detection attempts to identify densely connected subnetworks with relevant, potentially causal meaning. As searching the space of all communities is prohibitively expensive \cite{brandes2006maximizing}, heuristic-based approaches are developed, sourcing their principles from many fields of science, including physics, biology, sociology and other \cite{harenberg2014community,lancichinetti2009community,honghao2013community}.

Established community detection algorithms operate on a network's adjacency structure directly, and are thus specialized only for this task. Examples of such algorithms include the Louvain algorithm \cite{de2011generalized}, which maximizes the modularity score that approximates the network's connectivity patterns so that densely connected parts of the network remain grouped. Another well established algorithm is InfoMap \cite{rosvall2009map}, which operates using the ideas from the information theory. It encodes sampled random walks and attempts to find codewords, i.e. structures representing communities, so that their length is minimized. Intuitively, it samples the ``information flow'' across the network and maximizes such network partition, so that the flow remains captured in densely connected parts of a given network. Both InfoMap and Louvain algorithm scale to massive, real-world networks. The InfoMap also offers the insight into hierarchical community organization, which is also commonly present in real-world networks. Other recently introduced algorithms, which perform approximately the same as InfoMap and Louvain algorithm include, for example, Grothendieck's inequality communities \cite{guedon2016community} and SCORE \cite{jin2015fast}. We refer the reader to \cite{harenberg2014community} for a more extensive overview.

Quality of community detection can be evaluated with several measures, including ARI (Adjusted Rand Index) \cite{ARI}, NMI (Normalized Mutual Information) \cite{thomas1991elements} as well as Modularity Score \cite{clauset2004finding}. These measures, used in this work, are presented in Section~\ref{sec-CommunityQualityMeasures}.

\subsection{Semantic subgroup discovery}
\label{sec-SSDRelatedWork}

Semantic subgroup discovery (SSD) \cite{langohr2012contrasting,vavpetivc2013semantic} is a field of subgroup discovery, which uses ontologies as background knowledge in the subgroup discovery process. This methodology is capable of inducing rules from classification data, where class labels denote the groups for which descriptive rules are learned. In semantic subgroup discovery, ontologies are used to guide the rule learning process. For example, the Hedwig algorithm \cite{adhikari2016explaining, vavpetivc2013semantic} (used in this work) accepts as input a set of class-labeled training instances, one or several domain ontologies, and the mappings of instances to the relevant ontology terms. Rule learning is guided by the hierarchical relations between the considered ontology terms. Hedwig is capable of using an arbitrary ontology to identify latent relations explaining the discovered subgroups of instances. The result of the Hedwig algorithm are descriptions of target class instances as a set of rules of the form  TargetClass $\leftarrow$ Explanation, where the rule condition is a logical conjunction of terms from the ontology.
The details of the community-based semantic subgroup discovery approach \cite{vskrlj2019cbssd, cbssd_nfmcp} used in this work are provided in Section~\ref{sec-SSD}.

\section{Silhouette community detection}
\label{sec:scd}
In this section, we present the proposed method we named Silhouette Community Detection (SCD). We begin by describing the general setting and the rationale that led us to the proposed approach. We continue by describing the notions of network embedding, as well as the Silhouette score we adapt for the task of community evaluation. We finally present the formal description of the proposed approach along with the analysis of its computational and spatial complexity.

\subsection{Definitions}
This section serves as the introduction to the concepts used throughout this work. We first define the types of networks we consider, followed by definition of the community detection task.

\begin{definition}[Weighted network]
A weighted network $G$ is a tuple $(N,E,w)$, where $N$ is the set of nodes and $E$ is the set of edges defined as unordered node pairs. The weight function $w$ maps from edges to the field of real numbers, i.e. $w: E \rightarrow \mathbb{R}^+$, assigning a weight to each edge. \end{definition}

The proposed method also naturally handles unweighted networks, which can be understood as weighted networks where all weights are set to 1.

\begin{definition}[Network embedding]
Given a network $G=(N, E, w)$, a network embedding of $G$ is a mapping $f: G \rightarrow \mathbb{R}^{|N| \times d}$, where $N$ is the set of network nodes. The value $d$ is a parameter of the embedding referred to as the latent dimension of the output vector space.
\end{definition}

We do not explicitly define the properties of $f$, as the wealth of existing methods exploits various aspects of $G$. Note that when the value of the embedding dimension $d$ is 2, the embedding is a collection of pairs of real values, which can easily be visualized. The goal of network embedding methods is to maintain the relevant graph-topological properties in the obtained vector space as accurately as possible.

We continue by defining the notion of clustering, as used throughout this work.

\begin{definition}[Clustering]
A clustering function $c$ is a mapping  $c: \mathbb{R}^{|N| \times d}  \rightarrow \mathbb{N}^{|N|}$. Each element of the resulting vector represents the label of the cluster, assigned to the corresponding row of the input matrix. The number of clusters $k$ can be defined upfront, in which case numbers from $1$ to $k$ serve as labels for the clusters to which the input vectors belong.
\end{definition}

We conclude our list of definitions with the notion of community detection.

\begin{definition}[Community detection]
Let $G$ represent a network as defined above. Let $P(G)$ represent a partition of $G$ into $n$ non-overlapping subnetworks $\{G_1,G_2,\dots G_n\}$. Community detection refers to the process of finding $P(G)$ with respect to a quality function $q: P(G) \rightarrow \mathbb{R}$, such that the value of $q$ is maximized.
\end{definition}
Note that the definition of $q$  (community scoring function) was not explicitly stated, as existing community detection algorithms optimize for different $q$. Examples of well known $q$ include  modularity (Louvain algorithm) and average description code length (InfoMap).

\subsection{Network representation learning}
In this section, we discuss the two node embedding methods we used throughout the empirical evaluation. Both methods were re-implemented using efficient libraries for sparse matrix manipulation, which we also consider as added value of this work.

\subsubsection{Embedding by factorization}
\label{sec:rep}

Network embedding algorithms map input networks to dedicated vector spaces, where a node's neighborhood's properties are kept approximately intact. Since the considered networks do not contain any node or edge features apart from their weights, we consider shallow network embeddings. We first describe NetMF, a recently introduced network embedding methodology, which implements the embedding process as implicit matrix factorization \cite{qiu2018network}. For example, the well known DeepWalk algorithm \cite{perozzi2014deepwalk} was shown to approximate the following matrix:
\begin{equation*}
M_{DeepWalk} =  \log \bigg (\textrm{vol}(G) \bigg (\frac{1}{T}\sum_{r=1}^{T} (D^{-1} \cdot A )^{r} \bigg )\cdot D^{-1} \bigg ) -\log b;
\end{equation*}
\noindent where $T$ is the context window size, $b$ the number of negative samples, $D = diag(d_1,\dots,d_{|N|})$, where $d_i$ represents generalized degree of node $i$ and $A$ the adjacency matrix, and $vol(G)$ is the volume of a weighted graph, defined as: $\sum_{i,j} A_{i,j} $.
Such network embedding ideas originate from the initial word representation learner word2vec \cite{NIPS2013_5021}. We re-implemented NetMF in PyTorch, as the original version was implemented in now deprecated Theano library \cite{bergstra2010theano}. We refer the interested reader to the original paper for theoretical details regarding the method \cite{qiu2018network}.

\subsubsection{Embedding by personalized node ranking}
The other node embeddings we test are Personalized PageRank vectors, obtained by the Personalized PageRank with Shrinking algorithm, recently introduced as part of HINMINE methodology \cite{kralj2018hinmine}. Here, vectors representing personalized node ranks are computed using the power iteration discussed next, whose output consists of P-PR vectors. 

\begin{equation*}
  \gamma_{u}(i)^{(k+1)} = \alpha \cdot \sum_{j \rightarrow i}\frac{\gamma_{u}(j)^{(k)}}{d_{j}^{out}}+(1-\alpha) \cdot v_{u}(i);k = 1, 2,\dots
  \label{eqPR}
\end{equation*}

For each node $u \in V$, a feature vector $\gamma_u$ (with components $\gamma_u(i)$, $1\leq i \leq |N|$) is computed by calculating the stationary distribution of a random walk, starting at node $u$. The stationary distribution is approximated by using power iteration, where the $i$-th component $\gamma_{u}(i)^{(k)}$ of approximation $\gamma_u^{(k)}$ is computed in the $k+1$-st iteration as follows:

\noindent The number of iterations $k$ is increased until the stationary distribution converges to the stationary distribution vector (P-PR value for node $i$).
In the above equation, $\alpha$ is the damping factor that corresponds to the probability that a random walk follows a randomly chosen outgoing edge from the current node rather than restarting its walk. The summation index $j$ runs over all nodes of the network that have an outgoing connection toward $i$ (denoted as $j \to i$ in the sum), and $d_{j}^{out}$ is the out degree of node $d_{j}$. Term $v_{u}(i)$ is the restart distribution that corresponds to a vector of probabilities for a walker's return to the starting node $u$, i.e. $v_{u}(u) = 1$ and $v_u(i)=0$ for $i\neq u$. This vector guarantees that the walker will jump back to the starting node $u$ in case of restart\footnote{If the binary vector was composed exclusively of ones, the iteration would compute the global PageRank vector, and the considered equation would be to reduce to the standard PageRank iteration.} \cite{page1999pagerank}. The HINMINE version of this algorithm was additionally parallelized via Multiprocessing library\footnote{\url{https://docs.python.org/2/library/multiprocessing.html}} where speedups from 100\% to 400\% were observed.

\subsection{Cluster detection}
First, we formally state the clustering problem being solved. Let emb=$f(G)$ represent a computed $d$-dimensional node embeddings, thus emb $\in \mathbb{R}^{|N| \times d}$. Obtaining a node partition using a clustering algorithm (representing a mapping $c$) of choice can thus be stated as $$P(G) = c(f(G)).$$
In this work, we consider clustering using efficient miniBatch k-means algorithm \cite{sculley2010web}, which we briefly discuss next.
Given a set of row vectors $\mathcal{X} \subseteq \mathbb R^d$, the objective of miniBatch k-means is to find a set $C$ of $k$ cluster centers $C=\{c_i, \dots ,c_k\}\subseteq \mathbb{R}^{d} $ which minimizes the following sum:
\begin{equation*}
\sum_{r \in \mathcal{X}} \min_{c \in \mathcal{C}} \big ( \textrm{dist}(r,c) \big )
\end{equation*}
The dist in this work denotes the Euclidean distance ($\lVert r -c \rVert^{2}$), even though other distances can be used. This problem, albeit being NP-hard, can be approximated well using random cluster initializations. In this work, we exploit the k-means++ algorithm for the initialization step (see \cite{arthur2007k} for more details).
 The considered miniBatch k-means algorithm also leverages the notion of sparse cluster centres, inspired by the power law nature of word occurrences. Here, the idea is to emphasize the points which occur commonly, as the majority of the points (e.g., words) could be very sparsely distributed and thus contribute little to cluster assignment. One of the reasons we selected this variation of k-means as the clustering detection method is also the similar, heavy tailed nature of node connectivity, resembling that of word occurrences. This observation indicates that similar heuristics could perform well. The miniBatch k-means algorithm is thus used to extract $k$ clusters from the node embedding space.
We then discuss, how to detect whether the $k$ clusters represent a good partition.

\subsubsection{Estimating cluster quality with Silhouette score}
\label{sec-SilhouetteScore}

The Silhouette score was initially introduced in \cite{rousseeuw1987silhouettes}. The score was successfully used previously for development of novel categorical data clustering algorithms \cite{aranganayagi2007clustering}, as well as text clustering tasks \cite{hotho2002ontology}, and can be defined as follows. Assume the input data was clustered into $k$ distinct clusters. The average distance between a given point $i$ and the remainder of the cluster is computed as:
\begin{equation*}
a(i) = \frac{1}{|C_{i}|-1} \sum_{j\in C_i\setminus \{i\}}\textrm{dist}(i,j);
\end{equation*}
where the distance $\textrm{dist}$ is defined by the user. In this work, we employ the Euclidean distance, thus computing
$d(i,j) = \lVert i - j \rVert;$
The $C_i$ corresponds to the cluster the $i$ is part of. The second part of the Silhouette score estimates the dissimilarity with other clusters as follows:
\begin{equation*}
b(i) = \min_{i \neq j} \frac{1}{|C_j|}\sum_{j \in C_j}\textrm{dist}(i,j);
\end{equation*}
thus computing the smallest average distance of $i$ to the points of the cluster $C_j$. The Silhouette of a single point can be defined as:
\begin{equation*}
s(i) = \frac{b(i) - a(i)}{\max{[a(i),b(i)]}};
\end{equation*}
which holds if $C_i > 1$, and $s(i)=0$ if $C_i=1$. Note that the $s(i)$ falls in the interval $[-1,1]$. Intuitively, very low Silhouette values represent non-distinct clustering, values around 0 represent overlapping clusters and higher values represent more defined clusters.  Finally, estimating the global clustering translates to averaging the Silhouette score across the points of interest, as follows:
\begin{equation*}
\textrm{Silhouette}(P(G)) = \frac{1}{|N|}\sum_{i \in N} s(i);
\end{equation*}

Obtaining the Silhouette score for each node thus corresponds to the estimate of how well a given node is clustered, whereas averaging the scores across the considered partition $P(G)$ gives an estimate of the \emph{global} clustering quality score.

\subsubsection{SCD formulation}
We discussed first how the embedding space of nodes can be subject to $k$-means clustering, yielding potential node partition. Second, we showed how a given partition can be evaluated in terms of intra- and inter- cluster homogeneity via the Silhouette score. The missing part to be discussed in this section is the formal statement of the optimization problem at hand, as well as the numeric procedure used to derive the final $k$.

For readability purposes, we define as $\textrm{Silhouette}_{\textrm{G}}(k)$ the Silhouette score obtained using a given $k$ (parameter of the k-means algorithm). We thus assume the network node embeddings were obtained from $G$ before running the clustering algorithms.
The proposed Silhouette Community Detection (SCD) algorithm, summarized in Algorithm~\ref{algo:main} works as follows. 

  \begin{algorithm}[t]
  
   \caption{Silhouette Community detection (SCD)}
    \label{algo:main}
   \KwData{Network $G$}
   \Parameter{Stopping criterion $w$, set of embedding parameter sets $\mathcal{P}$, Embedding algorithm $f$, maximum number of considered clusters $K$, cluster evaluation interval $\gamma$}   
   globalQuality $\leftarrow$ $-\infty$ \Comment*[r]{Initialize split quality.}
   optimalSplit $\leftarrow$ $1$ \Comment*[r]{Initialize optimal split.}
   optimalPartition $\leftarrow$ $\emptyset$ \Comment*[r]{Initialize optimal partition.}
   validRange $\leftarrow$ generateValidRange($K$,$\gamma$)\Comment*[r]{Initial cluster range.}
   \For{$\textrm{parameterSet} \in \mathcal{P}$}{
     emb $\leftarrow$ $f(G, \textrm{parameterSet})$ \Comment*[r]{Embed nodes.}
     stoppingCriterion $\leftarrow$ $w$\;
     improved $\leftarrow$ False \;
     \While{$k \in$ validRange and stoppingCriterion $\neq 0$}{
       partition $\leftarrow$ MBKMeans($k$, emb)\Comment*[r]{Cluster.}
       quality $\leftarrow$ Silhouette(partition)\Comment*[r]{Evaluate.}
       stoppingCriterion $\leftarrow$ stoppingCriterion - 1\;
       \If{quality $>$ globalQuality}{
         imporoved $\leftarrow$ True\;
         globalQuality $\leftarrow$ quality\Comment*[r]{Update global quality.}
         optimalPartition $\leftarrow$ partition \Comment*[r]{Assign new optimum.}
         optimalSplit $\leftarrow$ $k$ \Comment*[r]{Assign optimum number of clusters.}
         stoppingCriterion $\leftarrow$ $w$ \Comment*[r]{Reset stopping criterion.}
       }
     }
     \If{Improved}{
       optimalPartition $\leftarrow$ fineGrained(optimalSplit, emb)\;
       globalQuality $\leftarrow$ Silhouette(optimalPartition) \Comment*[r]{Update global quality.}
     }
   }
   \Return{optimalPartition}\;
  \end{algorithm}
  
The algorithm traverses the space of embeddings of interest ($\mathcal P$). For each embedding, computed using an embedding procedure $f$, a parameter sweep across values of $k$ is conducted.
SCD employs a two-step approach to finding the optimal $k$. First, it traverses $k$ values
defined as part of the validRange---an interval of potential Silhouette optima.
This range of $k$ values is initially determined based on $K$, the maximum number of clusters to be considered, and $\gamma$, an interval of $k$ values being considered. We further demonstrate how $\gamma$ can be \emph{automatically} determined based on $K$ in Section~\ref{sec:formal}.
We define this interval as equally distributed natural numbers, where the distance between the numbers is uniformly distributed (e.g., we take every $10$-th number on the interval between 1 and 1{,}000).
  If the global Silhouette is improved during this parameter sweep (MBKMeans represents the miniBatch k-means algorithm and Silhouette the computation of a given partition's score), the $k$, as well as the exact partition are stored.
  
 The second step of finding the optimal $k$ is a fine-grained optimization step (fineGrained, line 22). Here, the neighborhood of the previously identified $k$ (elements of validRange) is explored in more detail---an interval around the $k$ is exhaustively inspected.
We additionally introduced a stoppingCriterion parameter, which stops the optimization, if Silhouette is not improved in $w$ iterations.
Once the SCD concludes, it yields the partition of the nodes (or elements of the vector space) into a finite set of communities.

To address the problem of finding the optimal $k$, we thus consider the following steps.
First, the space of possible $k$ values is not densely defined i.e. we test only every $n$-th $k$. Second, we introduce a stopping criterion---when no improvement is made for enough updates, the algorithm starts a fine-grained search around a currently optimal $k$ identified as part of the initial, coarse-grained $k$ sweep.

\subsection{Formal analysis}
\label{sec:formal}
In this section we overview and summarize the key parts of the proposed Silhouette community detection algorithm. We begin by formulating the optimization problem that is being solved, followed by the analysis of the relevant aspects of the computational complexity.

Let $\mathcal{P}$ represent the set of the node embedding parameters, $\mathcal{K}$ the set of candidate $k$ values representing the number of clusters. $\textrm{SilhouetteGlobal} (p,k)$ represents the graph $G$'s partition scored with a Silhouette score obtained when parameter set $p$ was used along with $k$ clusters to obtain communities. The proposed Silhouette community detection algorithm thus attempts to solve the following optimization problem:
\begin{align*}
P_{\textrm{final}}(G) = \argmax_{k \in \mathcal{K},p \in \mathcal{P}} \big [ \textrm{SilhouetteGlobal} (k,p) \big ].
\end{align*}
\noindent As the quality of the obtained communities depends on the clustering as well as the embedding algorithm, we then discuss the computational complexity of the two steps. Contemporary node embedding algorithms can perform in subquadratic time with respect to the number of nodes and have spatial complexity, which is linear with respect to the number of edges. The $k$-means clustering family of algorithms is quadratic in the worst case, yet, the miniBatch sparse version used in this work in practice performs very fast, as it takes the sparsity of the input space into account. Its complexity is $\mathcal{O}(\phi \cdot k \cdot |N| \cdot d)$ for a given $k$, where $\phi$ corresponds to the number of steps required by k-means++ initialization. The Silhouette computation can be performed in $\mathcal{O}(|N|^{2} \cdot d)$ time, indicating that the dimension of the embedding plays an important role in the performance of this final step. As one of the main bottlenecks of the proposed method, we recognize the number of cluster evaluations. Thus, the validRange method, discussed in Algorithm~\ref{algo:main}, can contribute notably to the execution time (values of $k$ considered). The total computational complexity of the approach is thus $\mathcal{O}(\phi \cdot k \cdot |N|^{2}\cdot d)$. As shown in the following sections, node embeddings are in practice computed less frequently than the clustering, rendering the method more sensitive to the $k$ parameter than to the embedding setting considered. Finally, as the dimensionality $d$ of the embedding can vary from as little as $5$ (hyperbolic embeddings \cite{nickel2017poincare}) to as much as $1000$ or more, we note that selecting the sufficient (and low-dimensional) network embedding can offer realizable speedups of several orders of magnitude.

Finally, we analytically derive an estimate for $\gamma$, the size of the $k$ sampling interval with respect to the maximum number of communities expected ($K$) as follows:
\begin{align*}
\gamma = \sqrt[3]{K^{2}}
\end{align*}
For readability purposes, we omit the derivation of this estimate to Appendix~\ref{appendix:gamma}. Note that the rationale behind introduction of this estimate is performance, as compared to the worst case, where $K$ different cluster sizes are considered, here we consider a substantially lower number and thus speed up the cluster detection process (this can result in an order of magnitude speedups).
In the following sections, we discuss the empirical setting, used to evaluate the performance of the proposed SCD algorithm.

\section{Experimental setting}
\label{sec:experimental}
In this section, we discuss the empirical evaluation used to assess the performance of the proposed approach. We first discuss the baseline network community detection methods, and continue with the description of the networks the methods were tested on.

\subsection{Considered algorithms}
We tested three existing community detection algorithms, described in Section~\ref{subs:baselines}, and two variants of SCD, described in Section~\ref{subs:scd}. 
\subsubsection{Baselines}
\label{subs:baselines}
We compare the proposed SCD approach against the following methods:
\begin{itemize}
\item \textbf{InfoMap} \cite{Rosvall2009}. This information flow-based algorithm represents a gold standard for community detection task.
\item \textbf{Louvain algorithm} \cite{clauset2004finding}. Similarly to InfoMap, Louvain algorithm is one of the most widely used community detection algorithms.
\item \textbf{Label propagation} \cite{cordasco2010community}. This simple baseline propagates the information in a breadth-first type of manner and serves as a weak baseline.
\end{itemize}

\subsubsection{SCD implementations tested}
\label{subs:scd}
We tested two implementations of SCD, based on representations, obtained by two network embedding algorithms; namely:
\begin{itemize}
\item \textbf{SCD - NetMF}. The Silhouette score optimization is conducted based on node representations obtained by the NetMF approach, which we re-wrote in PyTorch \cite{paszke2017automatic} for the purpose of this work.
\item \textbf{SCD - PPR}. The Personalized PageRank with Shrinking algorithm (PPR) is used to obtain stationary distributions of random walkers, representing a series of features for each node. The implementation is based on the one used by the HINMINE algorithm, introduced in \cite{kralj2018hinmine}. We used the version of the HINMINE algorithm which was further parallelized in \cite{vskrlj2018targeted}.
\end{itemize}

\subsection{Networks considered}
In this section, we discuss the networks we used for empirical evaluation of the proposed approach.

\subsubsection{Synthetic networks}
We conducted benchmarks over a space of synthetic Lancicinetti-Fortunato-Raddichi (LFR) networks \cite{lancichinetti2008benchmark}. This family of network models generates networks with corresponding ground truth communities. Such networks are commonly used to evaluate the community detection properties over a larger space of graphs with diverse topological properties \cite{lancichinetti2009community,yang2016comparative}.

The considered LFR networks are determined by the following parameters:

\begin{itemize}
\item Total number of nodes. 
\item Average node degree. 
\item Maximum node degree. 
\item Mixing. This parameter determines how well defined the generated communities are. It spans from 0 (very well defined) onwards, where, for example, graphs with mixing=1 have very poorly defined communities.
\item Degree exponent. Exponent of the node degree distribution (e.g., 2 implies power law network)
\item Community exponent. Exponent determining the community sizes.
\end{itemize}

We generated the space of networks defined by combinations of the following parameters: 
\begin{itemize}
\item Numbers of nodes:: [100,500,750,1000,2500,5000,10000]
\item Average node degrees: [15,30,50]
\item Maximum node degrees: [10,50,100,500]
\item Mixing: [0.1,0.2,0.5,0.7,0.9]
\item Degree exponent: 2
\item Community exponent: 1
\end{itemize}

\begin{figure*}[htb!]
\centering
\begin{tabular}{ccc}
\subcaptionbox{$|N|=1{,}000$}{\includegraphics[width=0.32\linewidth]{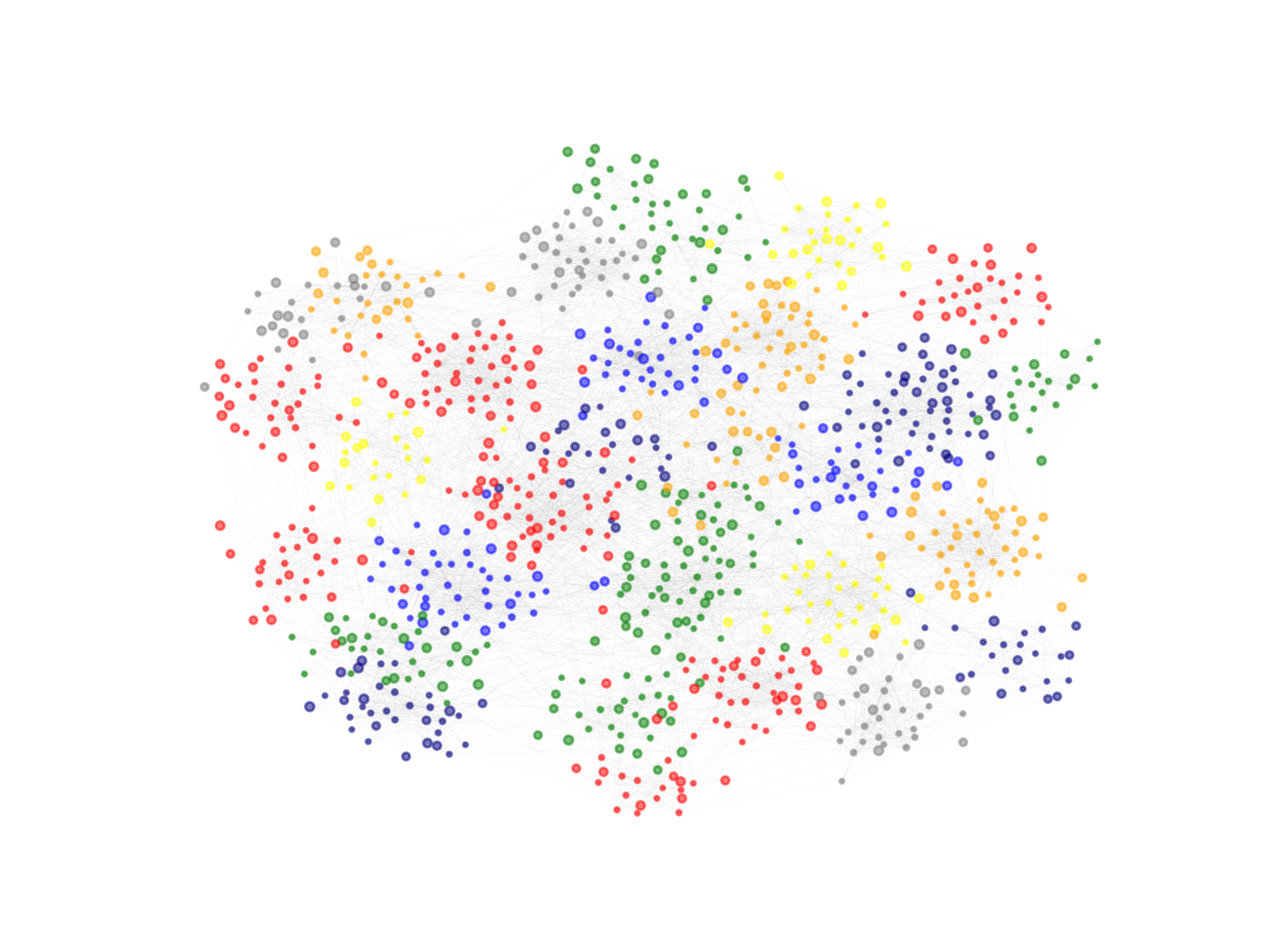}} &
\subcaptionbox{$|N|=5{,}000$}{\includegraphics[width=0.32\linewidth]{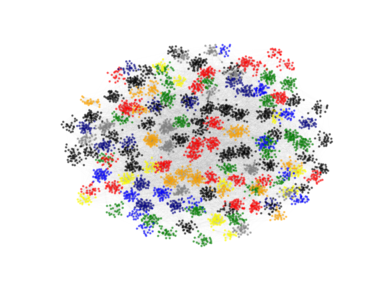}} &
\subcaptionbox{$|N|=10{,}000$}{\includegraphics[width=0.32\linewidth]{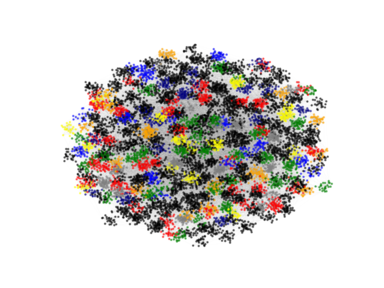}} \\
\end{tabular}
\caption{Three examples of LFR networks. The largest example LFR network consists of 10{,}000 nodes and 302{,}160 edges. The mixing parameter for these networks is set to 0.1, indicating very well defined communities. We colored the first 100 communities by size (random colors).}
\label{fig:lfr}
\end{figure*}

In total, we generated 234 valid networks with various topological properties. Example LFR network with highlighted communities is shown in Figure~\ref{fig:lfr}

\subsection{Real social network used}

\begin{figure}[ht]
\centering
\includegraphics[width=\linewidth]{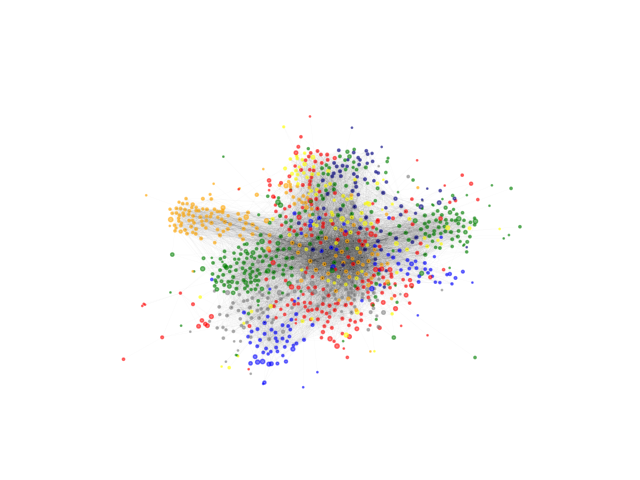}
\caption{E-mail ground truth communities (departments of senders).}
\label{fig:gt}
\end{figure}

Further, we test how well communities can be detected on a network with known ground-truth communities corresponding to E-mail network of one of the large European research institutions \cite{yin2017local}. An edge (u, v) in the network denotes that the person u sent person v at least one email. The e-mails only represent communication between institution members (the core), and the data set does not contain incoming messages from or outgoing messages to the rest of the world. The network consists of 1005 nodes and 25{,}571 edges, and is, along with its ground truth communities, visualized in Figure~\ref{fig:gt}.
The data set also contains "ground-truth" community memberships of the nodes. Each individual belongs to exactly one of the 42 departments at the research institute.

\subsection{Community quality evaluation measures}
\label{sec-CommunityQualityMeasures}

An established approach to evaluating the quality of community detection algorithms is via \emph{ground truth communities}. We assume the ``optimal'' partition $P(G)$ (ground truth) is known upfront. 
A partition similarity score (as defined next) is used to compare the ground truth partition with the one returned by a community detection algorithm.


We next discuss the measures of performance we used to evaluate how well a given algorithm is able to detect communities. We employ the following three measures.

\begin{description}
\item[\textbf{Normalized Mutual Information (NMI)}.] This index is defined as follows:
\begin{equation*}
\textrm{NMI}(Y,C) = \frac{2 \times I(Y;C)}{H(Y) + H(C)};
\end{equation*}
where $H(C)$ denotes the entropy of the assigned labels $C$, the $H(Y)$ entropy of the ground-truth labels $Y$. The $I(Y;C)$ denotes the mutual information \cite{thomas1991elements} between $Y$ and $C$. Thus, the larger the score, the better the matching.

\item[\textbf{Adjusted Rand Index (ARI).}]
This index measures a similarity between two clusterings by considering all pairs of samples and counting pairs that are assigned in the same or different clusters in the predicted and true clusterings. For readability purposes we do not define it here, yet we refer the reader to \cite{ARI} for the exact formulation.

\item[\textbf{Modularity.}]
The modularity measure \cite{clauset2004finding} is defined for a network partitioned into communities as follows:
\begin{equation*}
Q = \frac{1}{2m}\sum_{v=1}^n\sum_{w=1}^n \bigg [  A_{v,w} - \frac{k_{v}k_{w}}{2m} \bigg ]\delta(c_{v},c_{w})
\end{equation*}
\noindent where $n$ represents the number of nodes and $m$ the number of edges, $[A_{v,w}]_{v,w=1}^n$ denotes the adjacency matrix (i.e. $A_{v,w}$ is 1, when $u$ and $v$ are connected by an edge, and $0$ otherwise), $k_{v}$ denotes the degree of the $v$-th node and $c_v$ denotes the community the $v$-th node is assigned to. The $\delta(c_{v},c_{w})$ represents the Kr\"onecker delta function, which amounts to 1 when $c_v=c_w$ and 0 otherwise. The value $\frac{k_{v}k_{w}}{2m}$ represents the average fraction of edges between nodes $v$ and $w$ in a random graph with the same node degree distribution as the considered graph. Note that some of the baseline methods (e.g., Louvain algorithm) directly optimize the modularity, and are thus expected to perform favorably with respect to this metric. However, we believe computing modularity offers additional insights with respect to complementarity with other metrics, and potentially offers additional proof that the proposed Silhouette-based optimization indeed detects relevant signal.
\end{description}

\subsection{Other technical details}
\label{sec:tech}
In this section, we discuss some hardware-specific implementation details. The machine the benchmarks were run on was a Intel(R) Xeon(R) Gold 6150 CPU @ 2.70GHz processor equiped machine with 64GB of RAM. The machine also has a Nvidia Tesla GPU, which we used to test whether our NetMF implementation works as expected (on GPU). For the actual benchmarks, \emph{we did not use} GPU for factorizing the network, in order to more easily compare the execution times on CPU only. We intentionally didn't use GPU to demonstrate that no specialized hardware is needed to obtain competitive results. The LabelPropagation baseline was implemented using \cite{hagberg2008exploring}, the Louvain algorithm implementation, as well as a wrapper for the InfoMap binary can be found in \cite{py3plex}. The validRange for the empirical evaluation was set to the interval $[5,|N|,10]$. The embedding space parameters used during optimization were: number of negative samples ($\{1,5,20\}$), window size ($\{1,3,5,10,30,50\}$) and embedding dimension ($\{16,32,64,128,256\}$). In Appendix~\ref{appendix:tech}, we explain how the social network's Silhouettes were normalized based on the level of embedding dimension which yielded more robust results.

\section{Results---Community detection benchmark}
\label{sec:results}

In this section, we present the results of the empirical evaluation. We first discuss the results obtained on synthetic benchmark networks (Section~\ref{sec:synth}), followed by the results on the real-world social network (Section~\ref{sec:realw}).

\subsection{Results on synthetic networks}
\label{sec:synth}
Analyzing the 2-step search for the optimal value of $k$ in Algorithm \ref{algo:main}, we observe the proposed optimization in majority of cases finds a sufficient optimum---once the local optimum is found, some additional steps are performed to evaluate whether there exists a better solution in close range. In this work, we do not focus extensively on estimating the initial range of $k$, yet we believe such estimations could offer potentially better detection.

\begin{figure}[ht]
\centering
\includegraphics[width=\linewidth]{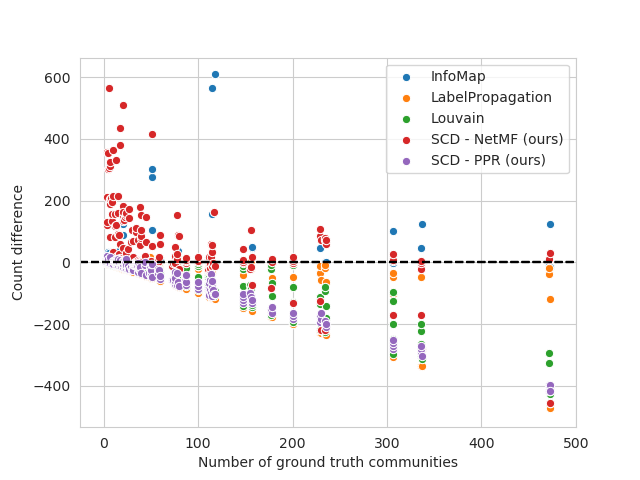}
\caption{Differences in the number of estimated communities (on LFR networks). The horizontal line represents optimal outcome with respect to the number of detected communities.}
\label{fig:count}
\end{figure}

The visualization of overall differences between the number of estimated communities and the number of ground-truth ones is shown in Figure~\ref{fig:count} --- the horizontal line represents the perfect match in the number of detected with that of ground truth communities. We can observe that the proposed SCD over-estimates the number of communities when small number of ground truth communities is present. However, the numbers stabilize when more than 100 communities are present. On the other hand, we can observe larger deviations with Louvain algorithm when the larger number of communities are present, indicating that Louvain algorithm algorithm under-estimates the number of communities. Similarly to SCD, InfoMap also over-estimates the number of communities when many communities are present, but for large numbers of ground truth communities, the over-estimation is more evident in InfoMap.

We believe performance with respect to the mixing parameter determining LFR graphs is of crucial importance, as it offers insight into how the considered community detection algorithms behave when communities are more or less defined. We show overall results, summarized with respect to this aspect in Table~\ref{tab:first_table}.

\begin{table*}[t]
\centering
\caption{Results over the space of synthetic networks. Values represent averages for each mixing level accross other parameters. We highlight top performers (first and second), with respect to NMI, as well as ARI scores. The LabelPropagation algorithm detects communities at higher mixing very poorly, we marked such runs with "*".}
\resizebox{0.9\textwidth}{!}{

\begin{tabular}{lllll} \hline
Mixing &           Algorithm &                  NMI &                  ARI &           Modularity \\ \hline
\rowcolor{green!15}   0.1 &             Infomap &  \textbf{0.994 $\pm$ 0.028} &   \textbf{0.996 $\pm$ 0.02} &  0.816 $\pm$ 0.118 \\
   0.1 &    LabelPropagation &  0.988 $\pm$ 0.038 &   0.96 $\pm$ 0.075 &   0.81 $\pm$ 0.133 \\
\rowcolor{green!5}   0.1 &             Louvain algorithm &   0.98 $\pm$ 0.031 &  0.925 $\pm$ 0.115 &  \textbf{0.817 $\pm$ 0.115} \\
   0.1 &  SCD - NetMF (ours) &   0.96 $\pm$ 0.089 &  0.924 $\pm$ 0.122 &  0.754 $\pm$ 0.195 \\
   0.1 &    SCD - PPR (ours) &  0.936 $\pm$ 0.051 &  0.867 $\pm$ 0.114 &  0.364 $\pm$ 0.307 \\ \hline
   0.2 &             Infomap &  0.922 $\pm$ 0.359 &  \textbf{0.939 $\pm$ 0.223} &  0.698 $\pm$ 0.181 \\
   0.2 &    LabelPropagation &  0.743 $\pm$ 0.961 &  0.832 $\pm$ 0.259 &  0.674 $\pm$ 0.217 \\
\rowcolor{green!15}   0.2 &             Louvain algorithm &  \textbf{0.969 $\pm$ 0.057} &  0.898 $\pm$ 0.136 &  \textbf{0.713 $\pm$ 0.127} \\
\rowcolor{green!5}   0.2 &  SCD - NetMF (ours) &   0.949 $\pm$ 0.11 &  0.911 $\pm$ 0.151 &  0.662 $\pm$ 0.174 \\
   0.2 &    SCD - PPR (ours) &   0.907 $\pm$ 0.09 &  0.815 $\pm$ 0.154 &  0.305 $\pm$ 0.269 \\ \hline
\rowcolor{green!5}   0.5 &             Infomap &  0.715 $\pm$ 0.903 &   \textbf{0.848 $\pm$ 0.32} &  0.411 $\pm$ 0.152 \\
   0.5 &    LabelPropagation &  0.378 $\pm$ 1.174 &    0.398 $\pm$ 0.3 &   0.326 $\pm$ 0.21 \\
   0.5 &             Louvain algorithm &  0.846 $\pm$ 0.238 &  0.657 $\pm$ 0.274 &  \textbf{0.435 $\pm$ 0.102} \\
\rowcolor{green!15}   0.5 &  SCD - NetMF (ours) &  \textbf{0.856 $\pm$ 0.197} &   0.75 $\pm$ 0.278 &  0.377 $\pm$ 0.137 \\
   0.5 &    SCD - PPR (ours) &   0.75 $\pm$ 0.233 &  0.502 $\pm$ 0.273 &  0.157 $\pm$ 0.179 \\ \hline
\rowcolor{green!15}   0.7 &             Infomap &  0.498 $\pm$ 1.135 &  \textbf{0.471 $\pm$ 0.467} &  0.166 $\pm$ 0.134 \\
   0.7 &    LabelPropagation (*) &  \textbf{0.679 $\pm$ 3.188} &      0.0 $\pm$ 0.0 &      0.0 $\pm$ 0.0 \\
\rowcolor{green!5}   0.7 &             Louvain algorithm &  0.648 $\pm$ 0.293 &  0.325 $\pm$ 0.241 &   \textbf{0.262 $\pm$ 0.04} \\
   0.7 &  SCD - NetMF (ours) &  0.492 $\pm$ 0.306 &  0.258 $\pm$ 0.367 &  0.118 $\pm$ 0.103 \\
   0.7 &    SCD - PPR (ours) &  0.407 $\pm$ 0.233 &  0.119 $\pm$ 0.195 &   0.03 $\pm$ 0.055 \\ \hline
\rowcolor{green!15}   0.9 &             Infomap &  \textbf{0.493 $\pm$ 2.529} &  0.001 $\pm$ 0.012 &  0.031 $\pm$ 0.051 \\
   0.9 &   LabelPropagation (*) &  0.744 $\pm$ 3.196 &      0.0 $\pm$ 0.0 &      0.0 $\pm$ 0.0 \\
   0.9 &             Louvain algorithm &  0.052 $\pm$ 0.024 &  0.002 $\pm$ 0.003 &  \textbf{0.164 $\pm$ 0.043} \\
   0.9 &  SCD - NetMF (ours) &  0.209 $\pm$ 0.109 &  0.003 $\pm$ 0.007 &   0.024 $\pm$ 0.03 \\
\rowcolor{green!5}   0.9 &    SCD - PPR (ours) &  0.193 $\pm$ 0.176 &  \textbf{0.019 $\pm$ 0.077} &  0.008 $\pm$ 0.024 \\ \hline
\end{tabular}
}

\label{tab:first_table}
\end{table*}

The results offer insights into performance of different algorithms with respect to various measures. As expected (and discussed in Section~\ref{sec:discussion}), the modularity score, which is optimized by the Loivain algorithm, is the highest with this algorithm. Other algorithms interchangeably outperform one another, indicating optimizing different metrics potentially leads to specialization in different parts of the networks space, thus leading to performance trade-offs.

\subsection{Results on a real network}
\label{sec:realw}
We also test the performance on a real world E-mail network with known ground truth communities.
\begin{table}[ht]
\centering
\caption{Community detection results on a real-world network representing E-mail senders. The LabelPropagation algorithm was not able to detect any communities, values of all scores were lower than $10^{-6}$.}
\begin{tabular}{c|ccc}\hline
  Algorithm &  NMI & ARI & Modularity \\ \hline
InfoMap & 0.654 & 0.301 & 0.39 \\
Louvain algorithm & 0.577 & 0.300 & \textbf{0.41}\\
LabelPropagation & $< 10^{-6}$ &  $< 10^{-6}$ &  $< 10^{-6}$  \\
SCD - PPR & 0.552 & 0.235 & 0.31 \\
 \rowcolor{green!15} SCD - NetMF ($d$ = 128) & \textbf{0.720} & 0.437 & 0.330 \\ 
  \rowcolor{green!15} SCD - NetMF ($d$ = 32) & 0.711 & \textbf{0.462} & 0.342 \\ 
 \hline
\end{tabular}
\label{tab:social}
\end{table}

\begin{figure*}[t]
\centering
\resizebox{1\textwidth}{!}{
\begin{tabular}{ccc}
\subcaptionbox{Ground truth}{\includegraphics[width = 1.3in]{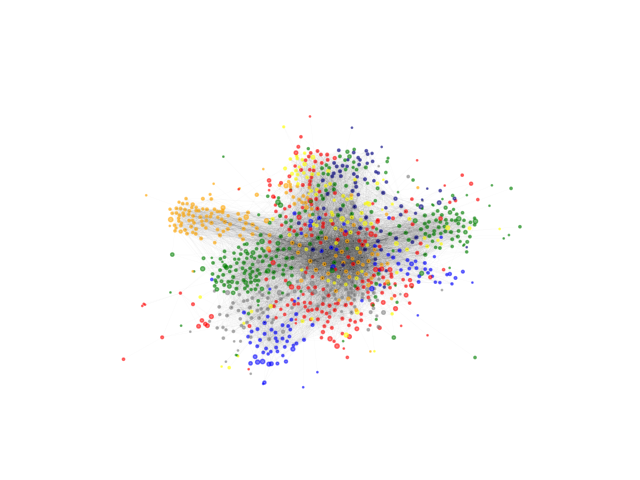}} &
\subcaptionbox{LabelPropagation}{\includegraphics[width = 1.3in]{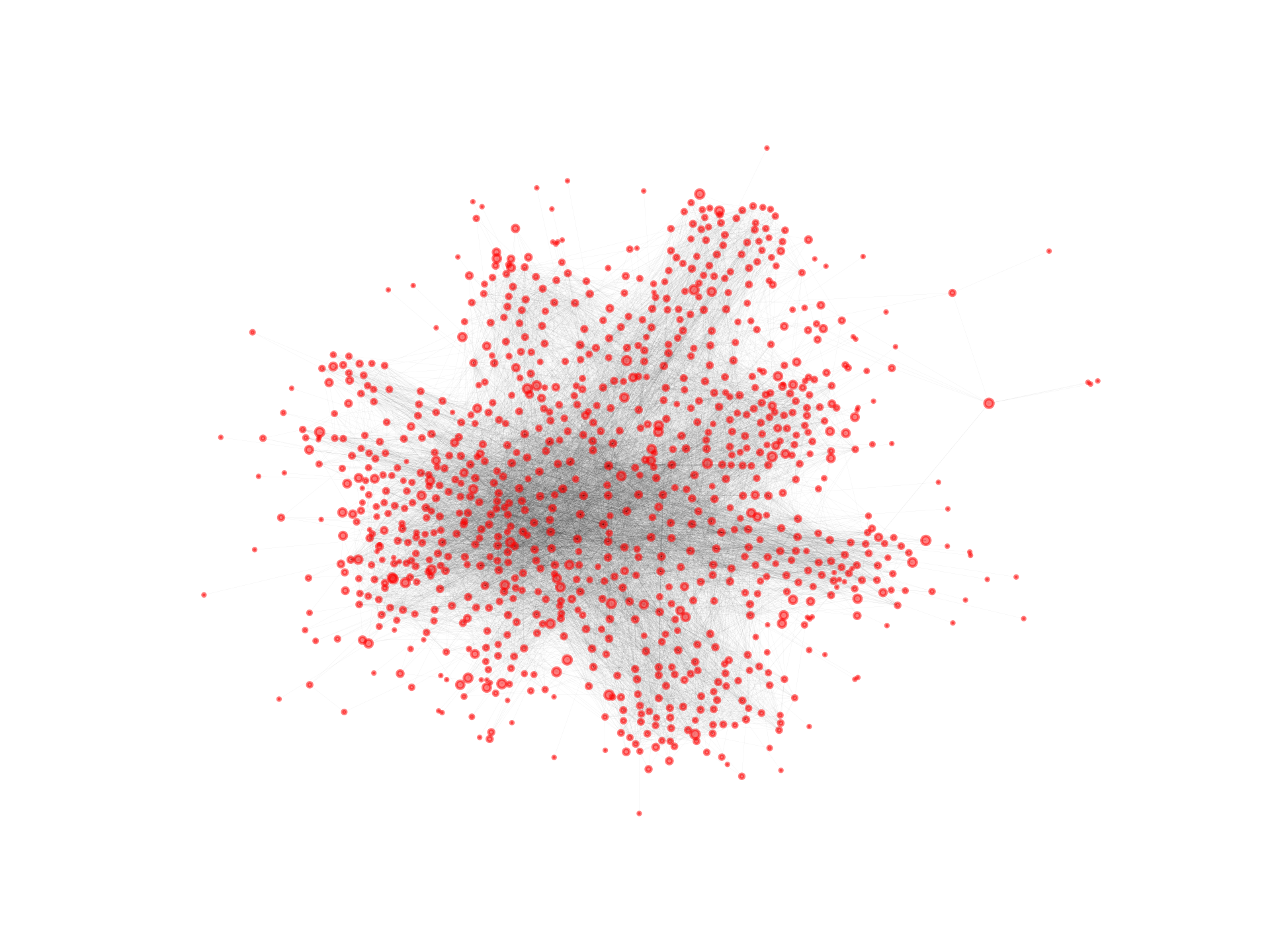}} &
\subcaptionbox{Louvain algorithm}{\includegraphics[width = 1.3in]{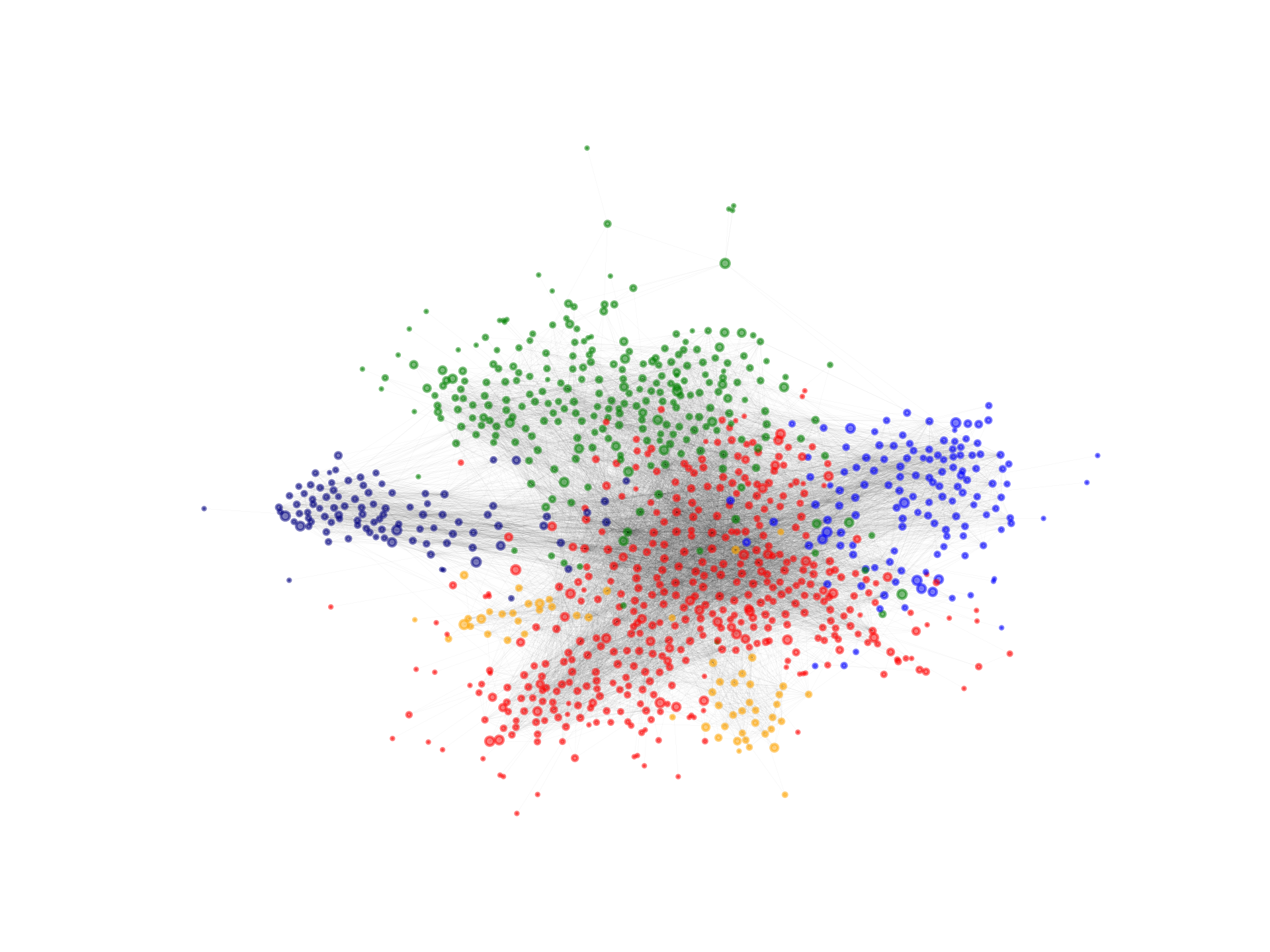}} \\
\subcaptionbox{InfoMap}{\includegraphics[width = 1.3in]{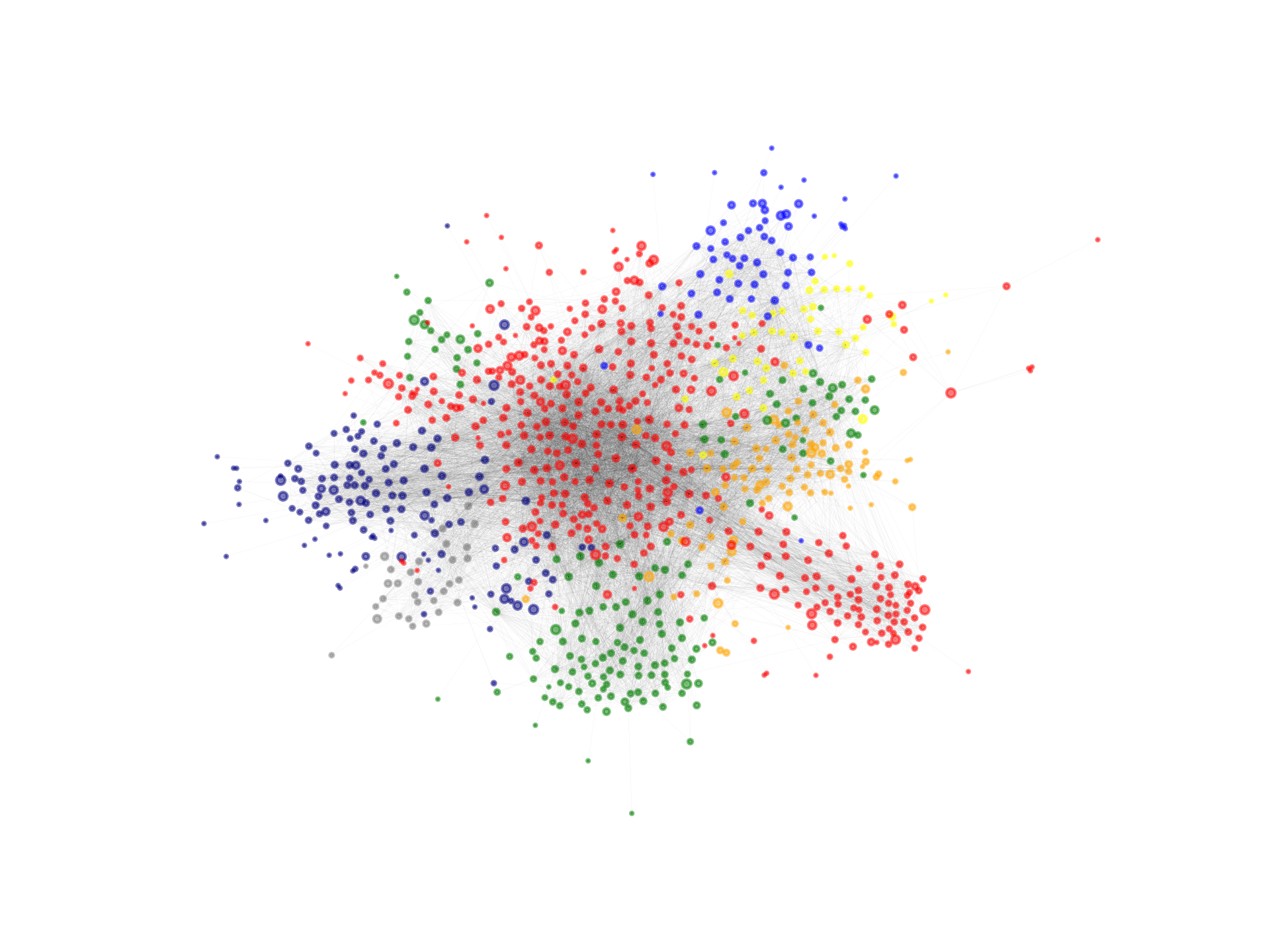}}&
\subcaptionbox{SCD - PPR}{\includegraphics[width = 1.3in]{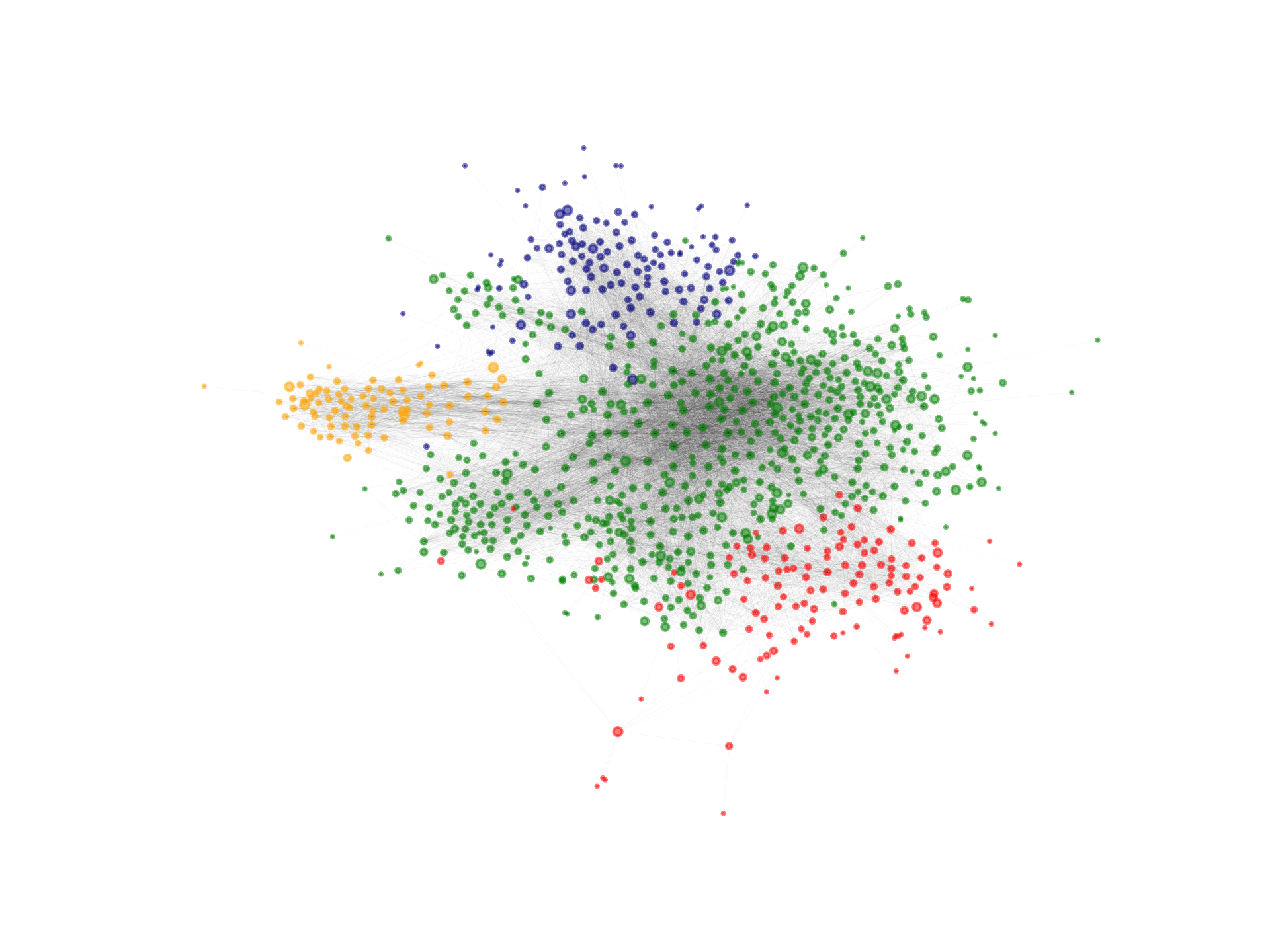}} &
\subcaptionbox{SCD - NetMF}{\includegraphics[width = 1.3in]{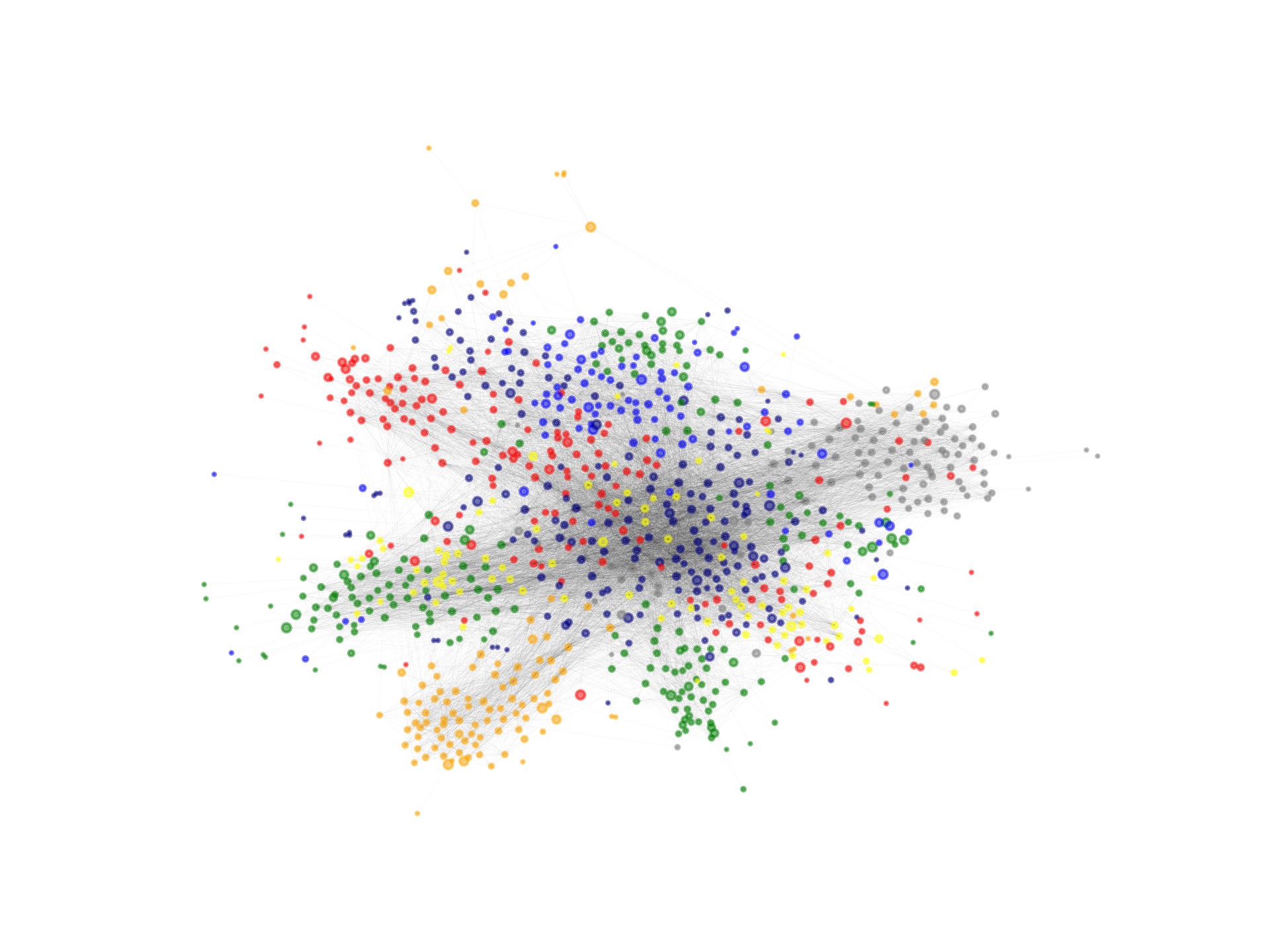}}
\end{tabular}
}
\caption{Visualization of E-mail network communities obtained using different algorithms. Communities are colored by size. The LabelPropagation and SCD - PPR performed the worst, which is also apparent from the visualizations---LabelPropagation did not detect any communities, whereas SCD - PPR detected too few.}
\label{fig:emb}
\end{figure*}

The proposed SCD approach  shows best performance on the mentioned real-life network (Table~\ref{tab:social}). 
The best performing embedding w.r.t NMI score was of dimension 128, with negative sampling parameter set to 1 and window size of 5. Interestingly, we also state the performance of a much smaller embedding ($d=32$), where the ARI score was better than in the case of the larger-dimensional one. Here, negative sampling was set to 1 and window size to 3. This result indicates that even embeddings of lower dimensionality potentially capture enough node similarity information that they are successfully grouped into communities.
We additionally visualize the results using the Py3plex \cite{py3plex} library in Figure~\ref{fig:emb}. The colors are based on community sizes---when communities are obtained (or given), they are sorted by size and colored with a predefined set of colors. Thus, the top three communities by size are colored red, green and blue. InfoMap, as well as SCD-NetMF ($d=128$) approaches detected the largest ground truth community, which is originally present in two parts (Figure~\ref{fig:emb}, subfigure a)), and as such harder to detect.
Similarly to the results obtained on synthetic networks, the Louvain algorithm, InfoMap and SCD - NetMF perform well---the communities they emit are visually distinct and resemble the ground truth network's ones. We can also observe SCD - PPR identified fewer communities compared to other approaches, indicating that taking whole stationary distributions into account is potentially not optimal. 

\begin{figure}[b]
\centering
\includegraphics[width=\linewidth]{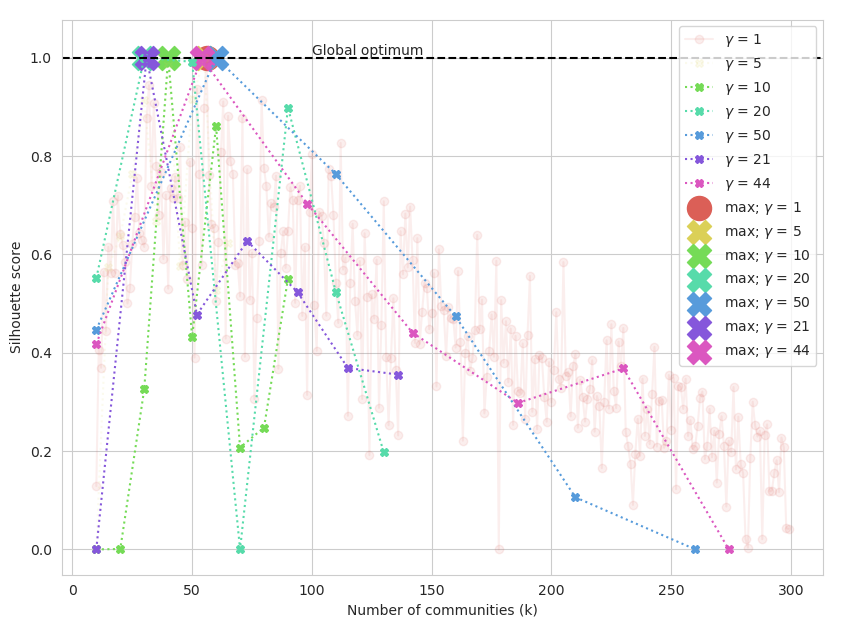}
\caption{Visualization of solutions found when different intervals of $k$ are considered. The situation where each $k$ is tested corresponds to $\gamma = 1$. The larger markers denote optima found when different intervals of $k$ are considered. It can be observed, all $\gamma$ variants yield similar number of communities ($\approx 50$), which is close to the ground truth of 42 communities.
}
\label{fig:brute}
\end{figure}

We next present the results of exhaustive empirical validation of the proposed Silhouette optimization procedure. For this task, we used the real-world social network discussed previously, and explored how the results of the proposed optimization correspond to a situation, where each $k$ is computed (exhaustive evaluation $\gamma=1$). The results for different values of $k$ are visualized in Figure~\ref{fig:brute}. 
Here, we normalized individual Silhouette scores obtained for different $\gamma$ values for readability purposes.
It can be observed that when different intervals of $k$ considered (denoted as $k$ in the Figure~\ref{fig:brute}) yield very similar results, where the number of communities is between 30 and 50 (number of ground truth communities is 42). This result indicates that the proposed SCD is not sensitive to $\gamma$, indicating potential speedups can be obtained, should this interval be selected based on e.g., a given network's properties---such situations are shown with pink ($K=300$) and violet ($K=100$), indicating the proposed $\gamma$ estimation offers good results. Note that individual SCD runs were for this figure run with stopping parameter set to 5---if after 5 iterations no improvement was made, the current optimum was returned as the result. The global Silhouette optimum (red points, indicating exhaustive search) was when $k=54$.

\section{Explaining communities with semantic subgroup discovery}
\label{sec:explain}
In this section we discuss how the obtained communities can be combined with background knowledge in the form of ontologies to provide human-understandable rules, describing individual communities. 


\subsection{Methodological background}
\label{sec-CBSSD background}
This section discusses how the in-house CBSSD methodology can be employed for obtaining such descriptions \cite{vskrlj2019cbssd}.
We first introduce the notions of subgroup and semantic subgroup discovery. Next, we discuss how communities can be understood as target classes for the task of semantic subgroup discovery. We also describe Hedwig, a semantic rule learner, which was for the purpose of this work parallelized to scale to thousands of candidate communities. 
In the following sections we discuss the key ideas of semantic subgroup discovery, as used in the remainder of this work.

\subsubsection{Subgroup discovery}
Subgroup discovery (SD) is a machine learning task where given a set of target classes and a set of instances, the goal is to identify significant patterns which describe a set of class-labeled instances (when compared to instances labeled differently). The goal of SD is thus similar to that of classification, with the main difference that in SD the emphasis is on individual patterns, which are symbolic, and thus explainable, whereas classification emphasizes construction of complete models (not necessarily explainable). The final result is thus not a predictive model but rather a series of interpretable rules, which serve to better understand the instance space.

\begin{definition}[Supervised learning]
More formally, given a set of classes $T$ and a set of class-labeled data instances $D$, the goal is to approximate the mapping $\Theta : D \rightarrow T$, which can explain/predict instances $d \in D$. 
\end{definition}
We next define rule learning as considered in this work.
\begin{definition}[Rule learning]
Let $\mathfrak{R}$ denote a set of all rules learned from  given data $D$ and class labels $T$. In rule learning, best rules $r_{1,\dots,n}\in \mathfrak{R}$ are found by optimizing a predefined success criterion, evaluated using a scoring function $\epsilon$,  $\epsilon: r_i \rightarrow \mathbb{R}$,  that assigns each identified rule $r_i$ a corresponding score in $\epsilon(r_i) \in \mathbb{R}$.
\end{definition}

In this work, we focus on subgroup discovery, a subfield of supervised descriptive rule induction \cite{novak2009supervised}. Here, a learner $\Theta$ is applied on a data set $D$ labelled with target classes from $D$. Similarly to supervised learning, $\Theta$ aims at identifying and describing interesting subsets of $D$ which are labelled with a given target class $t\in T$. Unlike supervised learning, where the result is a predictive model, the final result of descriptive learning are sets of rules. Each rule explains a subset of positive examples of selected class $t$. In general, the optimal set of rules is obtained by maximizing rule quality $\epsilon$.

\subsubsection{Semantic subgroup discovery} 
\label{sec-SSD}
Semantic subgroup discovery (SSD)
\cite{langohr2012contrasting,vavpetivc2013semantic} is a field of subgroup discovery that uses ontologies as background knowledge in the subgroup discovery process. The goal of SSD is to induce rules from labelled data, where (class) labels denote the groups for which descriptive rules are learned. For example, the Hedwig algorithm \cite{adhikari2016explaining, vavpetivc2013semantic} (used in this work) accepts as input a set of class-labeled training instances, one or several domain ontologies, and the mappings of instances to the relevant ontology terms.
Hedwig was successfully applied in the biomedical domain \cite{adhikari2016explaining}, supports RDF-encoded inputs, and is suitable for working with collections of background knowledge ontologies. Rule learning performed by Hedwig is guided by the hierarchical relations between the considered ontology terms. Hedwig is capable of using an arbitrary ontology to identify latent relations explaining the discovered subgroups of instances. The result of the Hedwig algorithm are descriptions of target class instances as a set of rules of the form  TargetClass~$\leftarrow$ Explanation, where the rule condition (Explanation) is a logical conjunction of terms from the ontology.

Class-labeled rules are usually learned via coverage-based approaches  \cite{furnkranz2012foundations}.
In this work we follow a different, recently introduced rule learning approach, which does not use a covering approach. In the selected approach implemented in the Hedwig algorithm \cite{vavpetivc2013semantic,vavpeticphd}, subgroup describing rules are learned using a specialized beam search procedure, and the output is a set of $b$ rules in the final beam of size $b$=$|Beam|$.

For an interested reader we here explain the formulation for rule induction used by the Hedwig algorithm. The presented formulation consists of two objectives; rule uniqueness and rule quality, which together form the joint scoring function as follows:
\begin{align}
  \mathfrak{R}_{opt} = \argmax_{\mathfrak{R}}\frac{\sum_{r \in \mathfrak{R}}\epsilon(r)}{\sum_{\substack{r_{i},r_{j}\in \mathfrak{R} \\ i \neq j}}|Cov(r_{i}) \cap Cov(r_{j})|+1}
  \label{eq:HedwigScore} 
\end{align}
\noindent where $\mathfrak{R}$ represents a candidate set of rules, $r \in \mathfrak{R}$ represents a single rule, and $Cov(r_i)$ denotes the set of examples covered by $r_i$. 

Hedwig aims to maximize the numerator of \ref{eq:HedwigScore} in order to maximize rule quality of a set of rules. At the same time, it searches for rules that cover different parts of the example space, which is achieved by minimizing the denominator, i.e. minimizing the intersection of instances covered by different rules $r_{i}$ and $r_{j}$.
In Hedwig, a set of rules (a beam of size $b$) is iteratively refined during the learning phase using a selected refinement heuristic, such as for example lift or weighted relative accuracy.
The algorithm yields multiple different rules that represent different subgroups of the data set being learned on.

\subsubsection{Community-based Semantic Subgroup Discovery (CBSSD)}

\begin{figure}
\centering
\includegraphics[width=\linewidth]{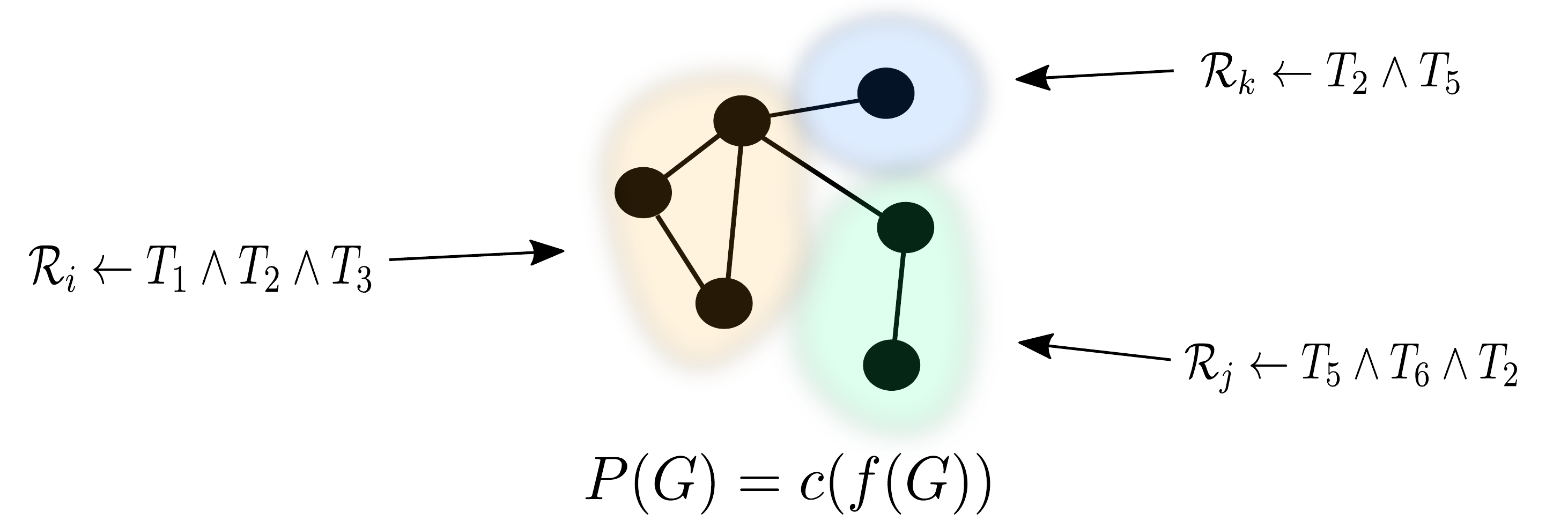}
\caption{Community-based Semantic Subgroup Discovery. Individual rules, such as $\mathcal{R}_{i,j,k}$ represent human-understandable descriptions, comprised of terms ($T_i$) of individual communities.}
\label{fig-CBSSD-new}
\end{figure}

Having defined the notions of rule learning and subgroup discovery, we next discuss (following \cite{vskrlj2019cbssd}) how Hedwig can be adapted for the task of \emph{community enrichment}---finding sets of rules which uniquely describe a given community. The CBSSD step, illustrated in Figure~\ref{fig-CBSSD-new}, can be understood as a post-hoc analysis to the proposed SCD. 
Note that this community explanation step is part of qualitative analysis, as the obtained patterns do not necessarily represent causal mechanisms---this part is commonly discussed with domain experts based on additional experimental evidence.
Thus, once communities are obtained using SCD, they can be enriched via ideas of CBSSD. We next describe some of the key ideas of CBSSD, yet direct an interested reader to \cite{vskrlj2019cbssd} for extensive technical details and computational complexity analysis.

Let $P_1 \dots P_n$ represent individual communities returned by SCD. Given background knowledge in the form of an ontology $\mathcal{B}$ and an injective mapping from a graph's nodes to such knowledge $m:N \rightarrow \mathcal{B}$, Hedwig is used as follows:
Each Partition is considered as a target class, whilst the remainder of the network is compared against. Thus, for $i$-th partition, rules of the form $$P_i \leftarrow t_1 \wedge t_2 \wedge \dots \wedge t_n $$ are learned, where $t_1$, $t_2 \dots t_n$ are elements of $\mathcal{B}$. For each community, a set of rules is learned, where the number of rules depends on the beam size used (input parameter).

\subsubsection{Speeding up Hedwig}
We next discuss the improvements we made to the original Hedwig algorithm for it to handle larger collections of background knowledge and hundreds of target classes (as required in this work) 
As the number of communities can be in the order of hundreds (or even thousands), one needs to consider $|P|$ (the number of communities) individual learners. This step can be time consuming (as shown in \cite{vskrlj2019cbssd}), thus we recognized that implementing Hedwig in parallel could introduce less time overhead. The implementation used in this work was parallelized at the class level. Here, we recognized as independent each class-specific learner. Thus, depending on the number of available CPU cores, Hedwig considers multiple communities simultaneously, offering from 5x to 15x speedups when compared to the original single CPU version.
\subsubsection{Parameters of the Hedwig semantic subgroup discovery algorithm}
\label{sec:pars}
We next
discuss how the task of semantic subgroup discovery was performed on obtained communities. For this task, we consider the Human Affinome, a collection of empirical protein interactions curated for the Affinomics consortium. The parsed graph consists of 1,171 nodes and 1,571 edges, with the average degree of 2.68. The same grid-based search for the maximum Silhouette as in the benchmark experiments (see Section~\ref{sec:tech})
was used. The following parameters were set for Hedwig---semantic subgroup discovery:
\begin{itemize}
\item The alpha value for determining rule significance was set to 0.05
\item FDR correction was used, where the threshold used was 0.1
\item Minimum support required was 0.01
\item Beam size of 30 was used
\item Depth was set to 10
\end{itemize}
The background knowledge considered was the whole gene ontology \cite{ashburner2000gene}, comprised, at the time of writing, of more than 40{,}000 terms.

As evaluation of rules for each community (separately) would be too time consuming, we selected the communities with the longest term conjuncts, as well as the most significant rules and performed literature-based evaluation of rules. 
As the discovered term conjuncts possibly represent well known biological interactions, we investigated individual rules separately, and compared their conjuncts to descriptions of genes, present in the studied community.

\subsection{Results}
\label{sec:results2}
\begin{figure}
\centering
\includegraphics[width=0.90\linewidth]{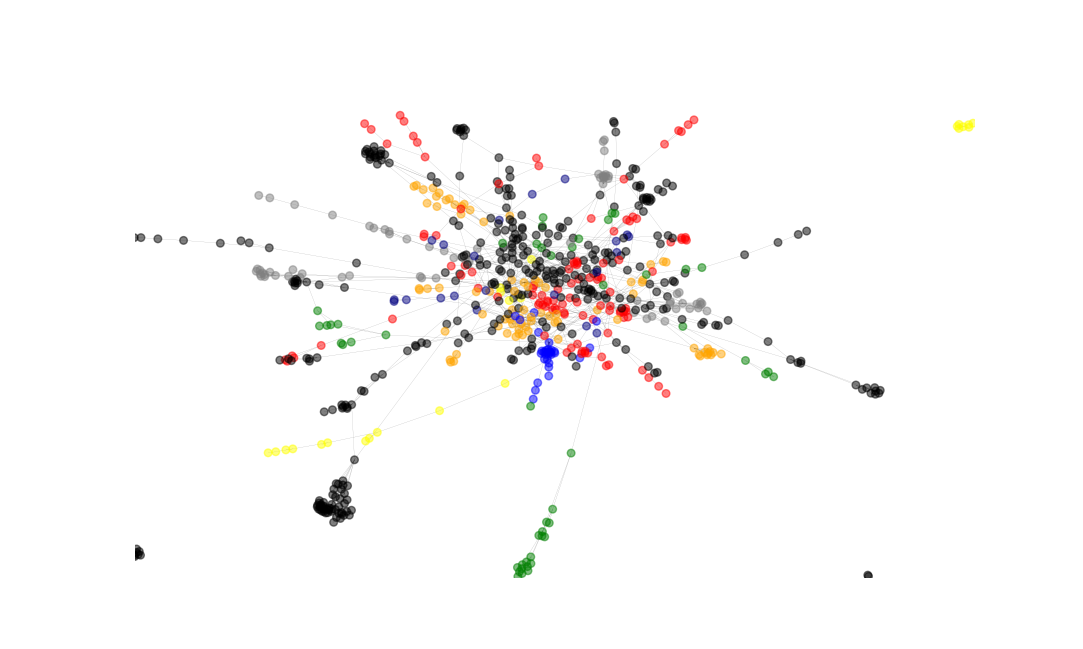}
\caption{Communities in the Human Affinome network}
\label{fig:affi}
\end{figure}

In Section~\ref{sec:pars}
we demonstrate the overall community detection performance of the proposed SCD. In this section we show the results of semantic subgroup discovery on the Affinome protein interaction network. This network was obtained based on extensive experimental evidence, and consists of interactions related to core metabolism.
The communities, detected in the Affinome are shown in Figure~\ref{fig:affi}.

\subsubsection{General overview of results}
The SCD algorithm discovered 145 communities. As the purpose of this work is not to explain every single one, but to demonstrate how they can be explained, we selected the community with the largest number of rules, as well as the larger number of multi-term conjuncts. The rules are summarized in Table~\ref{tbl:rules}.
\begin{table}
\centering
\caption{Results of subgroup discovery for the selected community. Section~\ref{sec:interpretation} discusses the meaning of the resulting GO terms with respect to the members of the resulting community.}
\begin{tabular}{ccccc}
\hline
                      Conjuncts & \#Positive &   Precision &    Lift &   Corrected p-value \\ \hline
                      GO:0007411 &   8 &  0.286 &  31.286 &  0.000 \\
                      GO:0038128 &   6 &  0.316 &  34.579 &  0.000 \\
                      GO:0007173 &   6 &  0.240 &  26.280 &  0.000 \\
 GO:0050900 $\wedge$ GO:0005515 &   6 &  0.188 &  20.531 &  0.000 \\
                      GO:0050900 &   6 &  0.167 &  18.250 &  0.000 \\
 GO:0038096 $\wedge$ GO:0005515 &   5 &  0.208 &  22.813 &  0.000 \\
                      GO:0038096 &   5 &  0.200 &  21.900 &  0.000 \\
                      GO:0001784 &   4 &  0.286 &  31.286 &  0.000 \\
                      GO:0008286 &   4 &  0.267 &  29.200 &  0.000 \\
                      GO:0038095 &   5 &  0.147 &  16.103 &  0.000 \\
                      GO:0005070 &   3 &  0.273 &  29.864 &  0.000 \\
 GO:0016303 $\wedge$ GO:0005515 &   3 &  0.250 &  27.375 &  0.000 \\
 GO:0008360 $\wedge$ GO:0005886 &   3 &  0.214 &  23.464 &  0.000 \\
                      GO:0016303 &   3 &  0.200 &  21.900 &  0.000 \\
 GO:0007169 $\wedge$ GO:0005524 &   3 &  0.188 &  20.531 &  0.000 \\
                      GO:0007265 &   3 &  0.150 &  16.425 &  0.001 \\
                      GO:0008360 &   3 &  0.143 &  15.643 &  0.001 \\
                      GO:0007169 &   3 &  0.143 &  15.643 &  0.001 \\
                      GO:0005884 &   2 &  0.200 &  21.900 &  0.003 \\
                      GO:0014065 &   2 &  0.200 &  21.900 &  0.003 \\
 GO:0031234 $\wedge$ GO:0004713 &   2 &  0.182 &  19.909 &  0.004 \\
                      GO:0071902 &   2 &  0.167 &  18.250 &  0.005 \\
                      GO:0097110 &   2 &  0.167 &  18.250 &  0.005 \\
 GO:0004715 $\wedge$ GO:0004713 &   2 &  0.167 &  18.250 &  0.005 \\
                      GO:0005901 &   2 &  0.143 &  15.643 &  0.006 \\
                      GO:0016337 &   2 &  0.143 &  15.643 &  0.006 \\
                      GO:0004715 &   2 &  0.143 &  15.643 &  0.006 \\
 GO:0042542 $\wedge$ GO:0005515 &   2 &  0.143 &  15.643 &  0.006 \\
                      GO:0031234 &   2 &  0.133 &  14.600 &  0.007 \\
                      GO:0042542 &   2 &  0.133 &  14.600 &  0.007 \\ \hline
\end{tabular}
\label{tbl:rules}
\end{table}

The members of the community the rules describe are summarized in Table~\ref{tbl:mems}. 

\begin{table}[ht]
\centering
\caption{Members of the described community.}
\begin{tabular}{ccl}
\hline
UniProt ID & Gene  & Name \\ \hline
P27986 & PIK3R1 & Phosphatidylinositol 3-kinase regulatory subunit alpha \\
P12931 & SRC & Proto-oncogene tyrosine-protein kinase Src \\
P46108 & CRK & Adapter molecule crk \\
Q07889 & SOS & Son of sevenless homolog 1 \\
P08631 & HCK & Tyrosine-protein kinase HCK \\
P42338 & PIK3CB & Phosphatidylinositol 4,5-bisphosphate 3-kinase \\
P29353 & SHC1 & SHC-transforming protein 1 \\
Q13480 & GAB1 & GRB2-associated-binding protein 1 \\
P01112 & HRAS & GTPase HRas \\
P07333 & CSF1R & Macrophage colony-stimulating factor 1 receptor\\ \hline
\end{tabular}
\label{tbl:mems}
\end{table}

\subsubsection{Interpretation of results}
\label{sec:interpretation}
This section interprets how the found rules are associated with the members of the enriched community. We systematically describe first what a given rule represents, followed by to what part of the community it maps.

We first observe, that many of the multi-conjunct rules contain the term ``GO:0005515'', which corresponds to ``protein binding''. The emergence of this very general term is expected, as the object of study is a protein interaction network. However, we notice that this term always appears in conjuncts with other terms, such as for example the ``GO:0042542'' (``response to hydrogen peroxide''), ``GO:0016303'' (``1-phosphatidylinositol-3-kinase activity''), ``GO:0038096'' (``Fc-gamma receptor signaling pathway involved in phagocytosis'') and ``GO:0050900'' (``leukocyte migration'').

We can observe that ``GO:0016303'', the term representing 3-kinase activity possibly emerged as the result of both PIK3R1, as well as PIK3CB proteins present in the studied community.

Next, both ``GO:0038096'', ``GO:0042542'', as well as ``GO:0050900'' represent events that are commonly present during immune response. We believe the aforementioned terms emerged as a consequence of CSF1R (Macrophage-colony factor receptor), HCK (Tyrosine-protein kinase), as well as SRC (Proto-oncogene tyrosine kinase). The tyrosine kinases transmit signals from cell surface receptors and play an important role in the regulation of innate \emph{immune responses}, including neutrophil, monocyte, macrophage and mast cell functions, phagocytosis, cell survival and proliferation, cell adhesion and migration. In combination with CSF1R the terms indicate the considered community is associated with immune response \cite{zhu2014csf1}.

We discuss the three most significant terms, namely ``GO:0007411'' (``axon guidance''), ``GO:0038128'' (``ERBB2 signaling pathway'') and 
``GO:0007173'' (``epidermal growth factor receptor signaling pathway''). The CRK protein, present in the considered community is known to regulate cell adhesion, spreading and migration. The ``GO:0038128'' represents a signaling pathway comprised of tyrosine kinases (present in the considered community). Similarly the association with the ``GO:0007173'' term related to epidermal growth factor signaling also corresponds to HRAS (GTPase HRas) and other kinases, which are known to play crucial roles during epidermal growth \cite{rosenberger2009oncogenic}.

To summarize, the proteins' functionality is indeed entailed in the obtained set of rules. Even though the considered community consists mostly of signaling and growth-related proteins, the related rules summarize key aspects such as cellular signaling and growth regulation, thus offering a human-interpretable description of the community without the time-consuming manual search.

\section{Discussion and conclusions}
\label{sec:discussion}
In this section we discuss the obtained results, as well as introduce potentially interesting further work.

\subsection{General overview}
We observe embedding-based community detection, as proposed in this work offers competitive performance on both synthetic, as well as real networks. One of the key observation is, SCD performs well if the embedding dimension is low (as can be observed from the computational complexity analysis). Thus, we recognize the recent achievements in the field of hyperbolic network embedding as potential further work. The proposed approach was tested with an efficient implementation of k-means clustering, yet, any clustering algorithm could be employed at this stage of community detection. Potentially more efficient alternatives could offer even faster performance.

Speed-wise, SCD - NetMF performed comparably to InfoMap, which was shown to scale to larger networks, even though the LabelPropagation and Louvain algorithm scaled even better. As discussed, the reduction of embedding dimension, as well as potentially less costly score, which is maximized could speed up the computation even further. However, we believe one of the benefits of the proposed SCD is that it can operate on pre-computed embeddings without any additional modification. This way, the complexity reduces to exploration of $k$ space, for which we theoretically, as well as empirically proved that it can be explored efficiently.

In terms of performance, the proposed method performs similarly to InfoMap and Louvain algorithm, optimizing a different measure of community quality (in a space where some information on the network structure is lost), potentially opening many new research venues. Even though we explored some parameterizations of the networks, we did not perform exhaustive search over the space of all embeddings, which we believe could potentially offer even better performance (at a significant computational cost).

The SCD also detected communities, which we interpreted using semantic subgroup discovery tool. Even though the aim of this analysis was not to discover novel knowledge, we were able to retrieve some existing (empiricaly proven) connections between proteins present in the same community, indicating such methodology could also offer novel knowledge when applied in a different setting.

\subsection{Further work}
We believe the conducted series of experiments also demonstrates, that modularity optimization is not necessarily the optimal tactic for finding the best partition. Modularity, although good at capturing densely connected parts of the graph, appears to miss-score the less apparent, but just as important connections.

Recent discoveries in the field of community number estimation also serve as complementary methodology to the proposed SCD. Here, the initial estimate of the number of communities could be notably improved (we employ a rather na\"ive scheme and do not consider any maximum-likelihood estimation).

\subsubsection{Non-euclidean geometries}
In this work, both the Silhouette computation, as well as the k-means computation were based on non-euclidean distance. Recent advancements in hyperbolic embeddings of real world networks offer novel insights into hierarchical organization underlying many such systems. Both k-means, as well as the Silhouette can be extended to hyperbolic spaces, for example the Poincar\'e disc, offering a natural extension for working with such non-euclidean embeddings.

\subsubsection{Complementarity with graph-convolutional networks}
Finally, we believe the proposed method is complementary to the recently emerging graph convolutional neural network embedding methodology. This branch of algorithms exploits features assigned to nodes (or edges) to obtain better node representations. As the resulting $\mathbb{R}^{|N| \times d}$ space is of the same type as considered in this work, we believe the proposed methodology could open an elegant extension to the research of community detection with meta information.

\subsubsection{Theoretical improvements}
Note that even though we offer a theoretical estimation of the $\gamma$ parameter, we believe the work could be further improved should e.g., maximum likelihood-based estimation of the number of communities in a given network be considered. Should such estimate be obtained, the space of $k$ values to be explored could be drastically reduced.

\subsubsection{Optimizing the embedding space separately}
In this work we performed rather na\"ive sweep through node embedding space in order to identify the configuration, which yielded the best Silhouette score. However, we believe that in certain applications, the node embeddings can be optimized w.r.t. a different task, e.g., classification, thus eliminating such expensive parameter search. We leave exploration of such claims for further work.

\subsubsection{Exploration of low-dimensional embeddings}
One of the interest results of this work is the fact that rather low dimensional embeddings (e.g. $d=32$) already yield good results in terms of community detection. We believe this aspect could be further explored, as $d$ is directly associated with computational complexity, thus reducing $d$ could yield multifold speedups, as well as offer insights into minimal dimension, needed to uncover a given network's latent structure.

\subsubsection{Feature-rich node embeddings as input}
The proposed SCD can cluster any (non-contextual) node embeddings. Thus, the recent body of work focusing on feature-rich networks, which yield real-valued vectors representing nodes can be naturally used with SCD for the task of community detection.

\subsubsection{Exploration of subnetwork clustering}
In this work we explored whether nodes can be grouped in a similar manner to that of contemporary community detection algorithms. However, instead of nodes, one can obtain embeddings of whole subnetworks. The proposed SCD can be naturally extended to such scenario, where, for example, very large networks could first be reduced to modular units, and the clustered. We believe this is one of the potential oportunities to scale SCD.

\section{Availability}
The SCD algorithm is freely available to academic users at \url{https://github.com/SkBlaz/SCD}

\section*{Acknowledgements}
The work of the first author was funded by the Slovenian Research Agency through a young researcher grant.
The work of other authors was supported by the Slovenian Research Agency (ARRS) core research programme \emph{Knowledge Technologies} (P2-0103) and ARRS funded research project
\emph{Semantic Data Mining for Linked Open Data} (financed under the ERC Complementary Scheme, N2-0078). The work was supported also by European Union’s Horizon 2020 research and  innovation programme under grant agreement No 825153, project EMBEDDIA (Cross-Lingual Embeddings for
Less-Represented Languages in European News Media).

\bibliographystyle{spmpsci}      
\bibliography{references}   

\appendix

\section{Closed-form solution for determining the $\gamma$ parameter}
\label{appendix:gamma}

In this section we present the derivation which led to an analytical estimate of the $\gamma$ (the interval in which the number of clusters are considered). We begin by theoretical analysis, followed by evaluation of how well the theoretical estimate fits to a large space of simulated networks.

\subsection{Problem statement}


The value of the optional parameter $\gamma$ in Algorithm~\ref{algo:main} presents a tradeoff between two extremes. On one hand, setting a low value of $\gamma$ means that each value $k$ from $1$ to $K$ must be checked. On the other, a high value of $\gamma$ means that 
the fineGrained step of SCD (see Algorithm~\ref{algo:main}, line 22), where a neighborhood of $k$, found in the first part of the Algorithm~\ref{algo:main} (lines 9-20) is exhaustively evaluated. For example, let the $k$ found in the first part (lines 9-20) be 10 and $\gamma=2$. The fineGrained method of SCD (line 22) additionaly explores the values of Silhouette when $k \in \{9,11\}$---a close neighborhood of the optimum found by the initial $k$ search.

To discover the optimal setting of $\gamma$, we first estimate the time complexity of Algorithm \ref{algo:main} when $\gamma=1$. This setting represents the baseline na\"ive sweep over all candidate values of $k$, and is compared to the search for $k$ with a given value $\gamma > 1$.

Without optimizing, with $\gamma=1$, Algorithm \ref{algo:main} does all of the work in its first step. The algorithm calculates $K$ Silhouette scores, each with a complexity of $\mathcal O(|N|^2d)$ and one $k$-means clustering for each $k\in[1..K]$. As the time complexity of the $k$-means clustering is $\mathcal O(k|N|d)$, the time complexity of this step is $\mathcal O(|N|d)+\mathcal O(2|N|d) +\cdots +\mathcal O(K|N|d) = \mathcal O(K^2|N|d)$. The entire iteration of Algorithm \ref{algo:main} therefore has a complexity of 

\begin{align*}\mathcal O(K\cdot |N|^2d)+\mathcal O(K^2|N|d) = \mathcal O(K|N|d(|N|+K))\end{align*}

Using a fixed value of $\gamma>1$, on the other hand, means that Algorithm \ref{algo:main} is composed of two phases. In the first phase, values of $k=\gamma, 2\gamma,\dots, K=\frac{K}{\gamma}\cdot \gamma$ are checked. This means that $\frac{K}{\gamma}$ runs of the Silhouette algorithm and $k$-means clustering for every checked value of $k$, for a total complexity of

\begin{align*}
\mathcal O\left(\frac{K}{\gamma}|N|^2d\right) + \mathcal O\left(\gamma|N|d+2\gamma|N|d+\cdots+\frac{K}{\gamma}|N|d\right)\\=\mathcal O\left(\frac{K}{\gamma}|N|^2d+\frac{|N|dK^2}{\gamma}\right) = \mathcal O\left(\frac{K|N|d}{\gamma}(|N|+K)\right)
\end{align*}

In the second phase, once the interval holding the optimal value of $k$ has been discovered, the $\gamma-1$ possible values of $k$ in that interval must still be checked. Assuming that the interval to check is $[n_0\gamma + 1.. (n_0+1)\gamma -1]$, this has a complexity of

\begin{align*}
(\gamma-1)|N|^2d + |N|d(n_0\gamma + 1 + n_0\gamma + 2+\cdots + n_0\gamma + \gamma - 1) \\= (\gamma -1)|N|^2d + \left(n_0+\frac12\right)(\gamma^2-\gamma)|N|d
\end{align*}
meaning that the total complexity of Algorithm \ref{algo:main}'s search for $k$ given a fixed value of $\gamma$ is

\begin{align*}
|N|d\left(\frac{K(K+|N|)}{\gamma}+|N|(\gamma - 1)+\left(n_0+\frac12\right)(\gamma^2-\gamma)\right).
\end{align*}

This means that the factor by which the time complexity is decreased by using $\gamma$ is equal to 

\begin{align*}
\frac{1}{\gamma}+\frac{n_0+\frac12}{K(K+|N|)}\alpha^2+\frac{|N|-n_0-\frac12}{K(K+|N|)}\alpha -\frac{|N|}{K(K+|N|)}.
\end{align*}

The optimal value $\gamma$ at which this factor is minimized can then be obtained as the solution to the equation

\begin{align}
- \frac{1}{\gamma^{2}} + 2 \cdot \frac{1}{\cdot K \cdot (K + |N|)} \cdot \gamma + \frac{k + |N| - 0.5}{K \cdot (K + |N|)} = 0.
\label{lab:eq}
\end{align}

While Equation~\ref{lab:eq} is solvable using Cardano's formulas, we are here only interested in the asymptotic behaviour of the optimal value of $\gamma$ in terms of varying values of $K$. Simplifying the solution of the equation shows, after some algebraic manipulation, that the optimal value of $\gamma$ is $\mathcal O(K^\frac{2}{3})$.

\subsection{Numerical evaluation of theoretical findings}
\begin{figure}
\centering
\includegraphics[width=\linewidth]{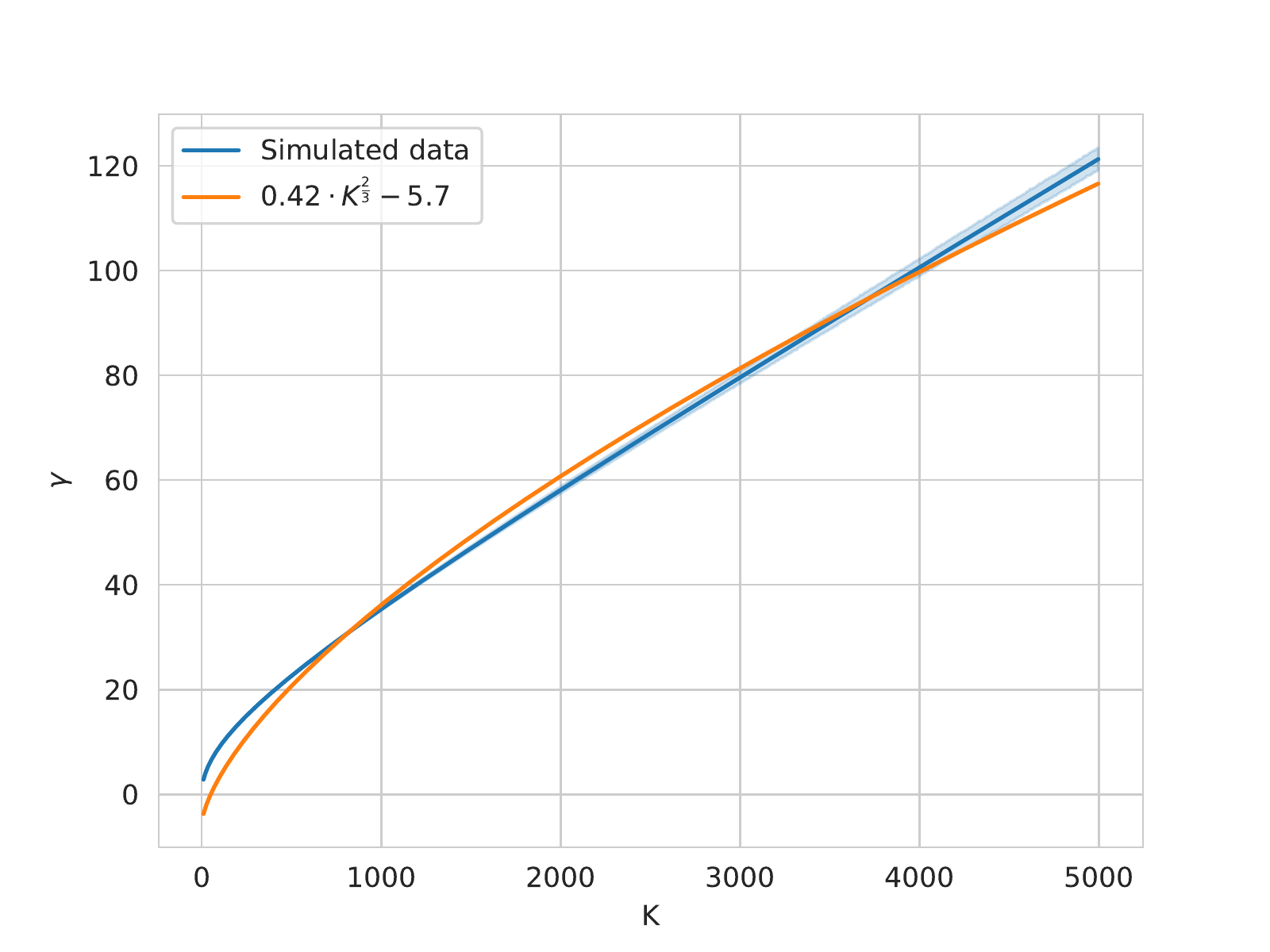}
\caption{Fitting the closed-form solution for determining the $\gamma$ parameter to simulated data.}
\label{fig:final}
\end{figure}

As discussed in the previous section, the connection between the parameter of interest $\gamma$ and parameters $K$---maximum number of communities considered, $k$, the actual number of communities and $|N|$, and the number of nodes in a given network is given in Equation~\ref{lab:eq}. 
Being analytically intractable, we simulated the solutions of this equation numerically, over the following parameter space: values of $K$, ranging from 10 to 5{,}000 in the increments of 10, $k$ ranging from 10 to 1{,}000 in the increments of 10 were considered. The number of nodes $|N|$ were [500,1000,2500,5000,10000]. For each parameter combination, we solved the Eq.~\ref{lab:eq} with starting conditions [10,50,100] and averaged the results (for stability purposes). The results are finally fit to the estimated relation between $k$ and $\gamma$, i.e.
\begin{align*}
\gamma = a \cdot \sqrt[3]{K^{2}} + b.
\end{align*}
The agreement between the simulated results and the assumed relation between $K$ and $\gamma$ is shown in Figure~\ref{fig:final}
It can be observed that the agreement between the orange line (model) and the blue (simulated data) is a reasonably good fit for most of the $K$ space.

\newpage
\section{Effects of embedding dimensionality on Silhouette}
\label{appendix:tech}
\begin{figure}
\centering
\includegraphics[width=0.8\linewidth]{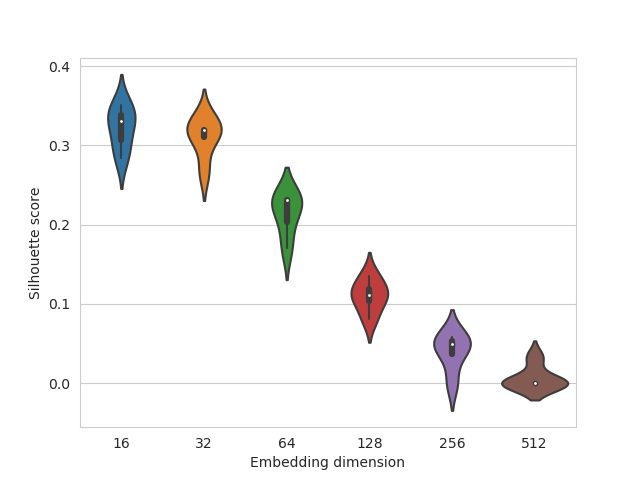}
\caption{Non-normalized Silhouette accross dimensions on the considered social network.}
\label{fig:non}
\end{figure}

\begin{figure}
\centering
\includegraphics[width=0.8\linewidth]{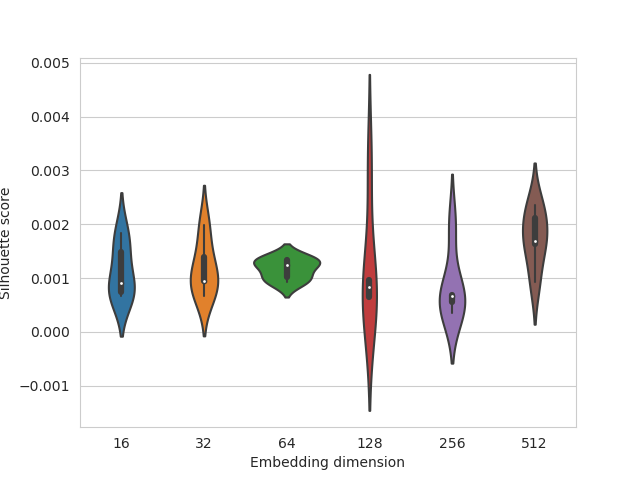}
\caption{Normalized Silhouette across dimensions on the considered social network.}
\label{fig:norm}
\end{figure}

In this section we offer evidence which led us to introduce optional embedding-level normalization of Silhouette score, yielding more robust community detection. The non-normalized scores are shown in Figure~\ref{fig:non}, whereas normalized ones are shown in Figure~\ref{fig:norm}. The plots show distributions of Silhouette scores across different embedding dimensions. Each distribution is based on grid search across the set of window sizes, as well as negative samples as discussed in Section~\ref{sec:tech}.

The normalized Silhouette scores are computed as follows. Let $\textrm{Silhouette}(P(G))$ represent the Silhouette score obtained for a given $k$ (number of clusters). Let $s_m$ represent the maximum Silhouette as identified using the implemented optimization and $s_i$ the $i$-th Silhouette considered when $k_i$ clusters were searched for. The normalized $s_m$, denoted $\mathfrak{s}_{m}$ is computed as:

\begin{align*}
\mathfrak{s}_{m} = \frac{s_m - \min{(s_i)}}{\max{(s_i)} - \min{(s_i})}
\end{align*}

Further, we observed that updating the best-performing embedding space based on mean normalized Silhouette values also yields more robust performance, yet evaluation of such claims is left for further work.
\end{document}